\documentclass[twocolumn,tighten,trackchanges]{aastex631}
\usepackage{amsmath,amssymb,amstext,bm}
\usepackage{graphicx, enumerate}
\usepackage{upgreek}

\submitjournal{ApJ}
\usepackage{float}
\usepackage{xcolor}
\usepackage{soul}

\defcitealias{2020ApJ...901...11H}{Paper I}
\defcitealias{1995ApJ...438..763G}{GS95}
\defcitealias{2000ApJ...537..720L}{LP00}
\defcitealias{2004ApJ...616..943L}{LP04}

\shorttitle{Radiative Transfer and the observed  Anisotropy in Turbulent Media}
\shortauthors{Hern\'andez-Padilla et al.}

\begin{document}

%%%%%%%%%%%%%%%%%%%%%%%%%%%%%%%%%%%%%%%%%%%%%%%%%%
\title{Effects of Radiative Transfer on the Observed  Anisotropy in MHD Turbulent Molecular Simulations}

\correspondingauthor{D. Hern\'andez-Padilla}
\email{david.hernandez@correo.nucleares.unam.mx}

\author[0000-0001-9574-1319]{D. Hern\'andez-Padilla}
\affiliation{
Instituto de Ciencias Nucleares, Universidad Nacional Aut\'onoma de M\'exico,
Apartado Postal 70-543, 04510 Ciudad de M\'{e}xico, M\'{e}xico}

\author[0000-0001-7222-1492]{A. Esquivel}
\affiliation{
Instituto de Ciencias Nucleares, Universidad Nacional Aut\'onoma de M\'exico,
Apartado Postal 70-543, 04510 Ciudad de M\'{e}xico, M\'{e}xico}

%\affiliation{Instituto de Astronom\'ia Te\'orica y %Experimental, CONICET - UNC, Laprida 854, X5000BGR %C\'ordoba, Argentina}

\author[0000-0002-7336-6674]{A. Lazarian}
\affiliation{
Astronomy Department, University of Wisconsin-Madison,
475 N. Charter Street, Madison, WI 53706, USA}
\affiliation{Centro de Investigaci\'on en Astronom\'ia, Universidad Bernardo O’Higgins, Santiago, General Gana 1760, 8370993, Chile}

\author[0000-0002-9644-1446]{P. F. Velázquez}
\affiliation{
Instituto de Ciencias Nucleares, Universidad Nacional Aut\'onoma de M\'exico,
Apartado Postal 70-543, 04510 Ciudad de M\'{e}xico, M\'{e}xico}

\affiliation{Instituto de Astronom\'{i}a y F\'{i}sica del Espacio (IAFE), Av. Int. G\"uiraldes 2620,
               Pabellón IAFE, Ciudad Universitaria, 1428, Buenos Aires, Argentina}

\author[0000-0003-1725-4376]{J. Cho}
\affiliation{
Department of Astronomy and Space Science, Chungnam
National University, Daejeon 305-764, Korea
}

%\watermark{1st Draft}

%%%%%%%%%%%%%%%%%%%%%%%%%%%%%%%%%%%%%%%%%%%%%%%%%%

\begin{abstract}

We study the anisotropy of centroid and integrated intensity maps with synthetic observations.
We perform post-process radiative transfer including the optically thick regime that was not covered in  \citet{2020ApJ...901...11H}. We consider the emission in various CO molecular lines, that range from optically thin to optically thick ($\mathrm{^{12}CO}$, $\mathrm{^{13}CO}$, $\mathrm{C^{18}O}$, and $\mathrm{C^{17}O}$). The results for the velocity centroids are similar to those in the optically thin case.  For instance, the anisotropy observed can be attributed to the Alfv\'en mode, which dominates over the slow and fast modes when the line of sight is at a high inclination with respect to the mean magnetic field. A few differences arise in the models with higher opacity, where some dependence on the sonic Mach number becomes evident.
In contrast to the optically thin case, maps of integrated intensity become more anisotropic in optically thick lines.
In this situation the scales probed are restricted, due to absorption, to smaller scales which are known to be more anisotropic. We discuss how the sonic Mach number can affect the latter results, with highly supersonic cases exhibiting a lower degree of anisotropy.

\end{abstract}


%%%%%%%%%%%%%%%%%%%%%%%%%%%%%%%%%%%%%%%%%%%%%%%%%%
%%%%%%%%%%%%%%%%% BODY OF PAPER %%%%%%%%%%%%%%%%%%%%%%%%

\section{Introduction}\label{sec:intro}

Molecular clouds are dense regions in the interstellar medium (ISM) that are largely affected by magnetic turbulence
\citep{1981MNRAS.194..809L, 2004ARA&A..42..211E,2007ARA&A..45..565M,2009ApJ...693.1074C}.
The study of its properties gives us information about, for instance, star formation \citep{2004RvMP...76..125M, 2007prpl.conf...63B},  acceleration and propagation of cosmic rays  \citep{2004ApJ...614..757Y}, heat transfer \citep{2001ApJ...562L.129N, 2006AIPC..874..301L}, and many other transport phenomena in the ISM \citep{2004ARA&A..42..211E}.

Since the 1950s, the ISM has been recognized as turbulent \citep[e.g.][]{1951ApJ...114..165V,1951ZA.....30...17V}.
Turbulence was invoked to explain observed line widths in spectra \citep{1981MNRAS.194..809L,1992MNRAS.256..641L,1984ApJ...277..556S,1987ASSL..134..349S}, velocity centroids \citep{1951ZA.....30...17V,1958RvMP...30.1035M,1985ApJ...295..479D,1985ApJ...295..466K,1994ApJ...429..645M}, or electron density fluctuations \citep{1989MNRAS.238..963N,1990ApJ...353L..29S}.

By the 1990s, there was evidence of a turbulent ISM observed at scales ranging from kiloparsecs to sub-astronomical units \citep{1995ApJ...443..209A,2010ApJ...710..853C}. The power spectrum of electron density fluctuations observed has a power law with a spectral index consistent with Kolmogorov type turbulence  \citep{1941DoSSR..30..301K}.
Nevertheless, characterizing the turbulence of velocity fields has been more difficult.

Later on, new techniques were developed to separate the contributions of the density and velocity fields, in order  to have reliable information on turbulence from radio spectroscopic observations.
\cite{2000ApJ...537..720L}, henceforth \citetalias{2000ApJ...537..720L}, develop a method called Velocity Channel Analysis (VCA) that is based on the analytical description of the emission in a position-position-velocity space (PPV). They assumed an optically thin medium and later extended to the optically thick case, absorption lines, and synchrotron radiation emission (\citealt{2004ApJ...616..943L}, henceforth \citetalias{2004ApJ...616..943L}, \citealt{2006ApJ...652.1348L}, see also \citealt{2009RMxAC..36...54P}). This analytical description has allowed the interpretation of observational data including, for example, \ion{H}{1} \citep{2001ApJ...551L..53S}, $\mathrm{^{13}CO}$ \citep{1998A&A...336..697S,2006MNRAS.372L..33B}, $\mathrm{C^{18}O}$ \citep{2006ApJ...653L.125P}.

In addition to the theory-based VCA, empirical techniques have been developed to obtain the spectral indices of density and velocity. For instance, the Spectral Structure Function \citep[SCF,][]{1999ApJ...524..887R,2001ApJ...547..862P} approach applies structure functions rather than spectrum calculations while obtaining the statistics of channel maps. The application of the Principal Component Analysis to spectral data  \citep[PCA,][]{1997ApJ...475..173H,2002ApJ...566..289B} also provides an insight into the statistics of the velocity, especially in the case when velocity fluctuations arise from localized clouds. As a new interesting development, the fluctuations of interstellar gas velocity were found to be reflected in the statistics the velocities of the newly formed stars \citep{2021ApJ...910...88X}, providing yet another way to study velocity turbulence.

For the VCA, the decontamination of the velocity statistics by density fluctuations can be achieved both choosing thin channel maps \citepalias[see][]{2000ApJ...537..720L} and using the Velocity Decomposition Algorithm, henceforth VDA, \citep{2021ApJ...910..161Y}. The latter is based on the \citetalias{2000ApJ...537..720L} description of intensity fluctuations in PPV space, but is focused on point-wise separating contributions of velocity and density contributions.
%With the help of the VDA, in \citet{2022arXiv220207871Y} for the first time the Big Power Law of Velocity fluctuations has been discovered.
It is important, that unlike the earlier turbulent power laws, the velocity turbulence spectrum continuously protrudes through various phases of the ISM that have very different densities. This suggests that the interstellar turbulence is, indeed, a universal process that governs both the evolution of the ISM and that of star formation.

We know that the ISM turbulence, in the presence of magnetic fields, is anisotropic. The anisotropy is described by \citet[][hereafter \citetalias{1995ApJ...438..763G}]{1995ApJ...438..763G} and is present in terms of the local direction of magnetic field \citep{1999ApJ...517..700L,2000ApJ...539..273C}. The consequence of this is that the in the system of reference related to the mean magnetic field, i.e. in the system of reference of the external observer, the anisotropy is determined by the anisotropy of the largest eddies \citep{2000ApJ...539..273C, 2001ApJ...554.1175M, 2002ApJ...564..291C}. The anisotropy of eddies at smaller scales is more prominent, but not usually seen after averaging along the line of sight is performed.
%They argue that the eddies are strongly anisotropic; that is, the eddies elongate in the direction of the magnetic field lines. This effect is growing on smaller scales.

 In the 1930s, molecules were already detected in space by optical absorption studies \citep{1937ApJ....86..483S}. By the 1960s, OH was detected in microwave spectra \citep{1963Natur.200..829W}, followed by the detection of $\mathrm{NH_3}$ \citep{1968PhRvL..21.1701C}, $\mathrm{H_2O}$ \citep{1969Natur.221..626C}, and formaldehyde \citep{1969PhRvL..22..679S}.
 Soon, the detection of molecular Hydrogen $\mathrm{H_2}$ in the UV part of the spectrum \citep{1970ApJ...161L..81C} and of carbon monoxide $\mathrm{CO}$ at $2.6$ mm \citep{1970ApJ...161L..43W} opened a new era of investigation of the molecular clouds.
 Empirical studies indicated that the $\mathrm{CO}$ to $\mathrm{H_2}$ ratio was close to $\sim 10^{-4}$ in dense molecular clouds. Molecular isotopologues $\mathrm{^{12}CO}$, $\mathrm{^{13}CO}$, and $\mathrm{C^{18}O}$ are regarded as tracers of $\mathrm{H_2}$ with numerical densities of $10^2$ and $10^4$ cm$^{-3}$, which correspond to  typical densities of young self-gravitating molecular clouds \citep{2012ARA&A..50...29C}. Since their lines tend to have lower optical depths, the $\mathrm{^{13}CO}$ and $\mathrm{C^{18}O}$ isotopologues can trace molecular gas over a wide range of densities. In contrast, the optically thicker $\mathrm{^{12}CO}$ isotopologue is used to trace lower-density outer regions of clouds.

\citetalias{2004ApJ...616..943L} predicted that the slope of the power spectrum of integrated intensity maps of an optically thick medium saturates at $-3$. Many observations are consistent with this prediction, but it was not until \citet{2013ApJ...771..123B,2013ApJ...771..122B} that these predictions were confirmed using numerical simulations with radiative transfer effects that simulated the $\mathrm{^{13}CO}$ $\mathcal{J}= 2\xrightarrow{}1$ transition.

\citet{2014ApJ...790..130B} expanded on the method of velocity centroids of \citet{2005ApJ...631..320E} to estimate the media magnetization, i.e. the ratio of turbulent to magnetic energies, with simulations that, among other observable characteristics, have the emission of lines of the $\mathrm{^{13}CO}$ $\mathcal{J}= 2\xrightarrow{}1$ transition. They found that for sub-Alfv\'enic turbulence, the $\mathrm{CO}$ emission shows a considerable anisotropy in the velocity centroid maps of $\mathrm{CO}$. At the same time, for super-Alfv\'enic turbulence it remains isotopic.

\citet{2015MNRAS.446.3777B} analyzed chemical models to have better insight into how the choice of chemical species as gas tracers influences the two-point velocity centroid statistics using  $\mathrm{^{12}CO}$ and $\mathrm{^{13}CO}$ molecules in the $\mathcal{J}= 1\xrightarrow{}0$ transition. They reported of significant consequences of changing optical depth while analyzing structures of the centroid velocity increments. Their power spectrum slope variations were in good agreement with the result theoretically predicted in \citetalias{2004ApJ...616..943L}.

Recently, \citet{2021ApJ...910...88X} studied the anisotropy of Faraday rotation and velocity centroids of the structure-functions  of the mean magnetic field using supersonic and sub-Alfv\'enic MHD simulations. They generate synthetic observations of $\mathrm{^{12}CO}$ and $\mathrm{C^{18}O}$ emission lines by calculating the radiative transport to their simulations. They find that the anisotropy obtained with the low-density tracer $\mathrm{^{12}CO}$ has a consistent dependence with $\mathcal{M}_A^{-4/3}$ (which is a measure of the isotropy degree). Anisotropy measured with the higher density tracer $\mathrm{C^{18}O}$ is lower and has a weak dependence on $\mathcal{M}_A$.

\citet{2021_unknown} extended the study of Structure-Function Analysis \citep[SFA described in][]{2021ApJ...911...37H,2021ApJ...910...88X} to measure the orientation and the $\mathcal{M}_A$ of the 3D magnetic field. Following \citet{2016MNRAS.461.1227K}, they confirm that the anisotropy observed in the intensity structures in the PPV space is regulated by the width of the velocity channel, the viewing angle $\gamma$ (between the direction and the LOS of the 3D magnetic field), and by the of Alfv\'enic Mach number.

In \citet[][hereafter \citetalias{2020ApJ...901...11H}]{2020ApJ...901...11H}, we use a grid of simulations similar to those in \citet{2015ApJ...814...77E}. We update those simulations to a uniform resolution of $512^3$ cells,  to produce synthetic spectroscopic observations (PPV data), and study the anisotropy in the structure-function of velocity centroids maps. With help of the procedure laid out in \citet{2002PhRvL..88x5001C,2003MNRAS.345..325C}, we decomposed the original velocity field into each of the MHD modes (Alfvén, slow, and fast MHD modes).
We studied how each mode contributes to the observed anisotropy as a function of the angle  between the line of sight and the mean magnetic field ($\gamma$). When the angle is large, the Alfvén mode dominates the observed anisotropy. While for smaller angles, the statistics are dominated by the slow mode. We compare our results with the analytical predictions in \citet{2016MNRAS.461.1227K,2017MNRAS.464.3617K}, and found that they are in reasonably good agreement, recovering most of the general trends and the level of anisotropy for the various models. Also, we study the integrated intensity maps. They show some anisotropy but not as pronounced as that observed in maps of velocity centroids.

In this work, we use the MHD simulations to simulate molecular lines maps of CO isotopologues (namely $\mathrm{^{12}CO}$, $\mathrm{^{13}CO}$, $\mathrm{C^{18}O}$, and $\mathrm{C^{17}O}$) in the $\mathcal{J} = 1\xrightarrow{}0$ transition. We generate molecular lines maps using a radiation transfer post-processing code based on that presented in \citet{2018MNRAS.478.2056T}.

We review the method for measuring velocity centroids anisotropy in observations in Section \ref{sec:centroids}. In Section \ref{sec:models} we describe our set of MHD numerical simulations and outline the main points of the radiative transfer algorithm. The results of our study, namely, the dependence of integrated intensity with optical depth and the isotropy degree as a function of the Alfvénic Mach number obtained for each vision angle and each optical depth, are presented in Section \ref{sec:results}, the effect of turbulence in the depth probed by the observations is addressed in \ref{sec:depth}. Lastly, in Section \ref{sec:disc}, we discuss our results, followed by a summary in Section \ref{sec:sum}.
%%%%%%%%%%%%%%%%%%%%%%%%%%%%%%%%%%%%%%%%%%%%%%%%%%

\section{Turbulence statistics from PPV data}\label{sec:centroids}

In \citetalias{2020ApJ...901...11H} we have studied the anisotropy of the structure-functions in synthetic observations from a grid of MHD turbulence simulations. To construct the synthetic observations (PPV data) we varied the line of sight, and assumed that the emission was optically thin.
In the present work, we extend our previous analysis to account for the self-absorption in different molecular lines that range from optically thin to optically thick.
We do this by considering the radiative transfer of those molecular lines in a post-processing step. The result of this step is a PPV data cube, which can be treated as observations: for instance, we can get 2D maps of integrated intensity and velocity centroids. To have different LOS, we rotate the output from the simulations by an angle $\gamma$ around the $y$-axis so that $\gamma$ corresponds to the angle between the LOS and the mean magnetic field (which is aligned with the original $x$-axis). Naturally, the  radiation transfer has to be computed for each model, molecular line, velocity mode (see Section \ref{sec:models}), and value of $\gamma$.

From the PPV data, we can obtain the integrated intensity as
\begin{equation}
    \label{eq:Int}
    I_\mathrm{\gamma}\left(\bm{X} \right)=\int \mathcal{I}_{\gamma}\left(\bm{X},v_\mathrm{los} \right)\,dv_\mathrm{los},
\end{equation}
where $\mathcal{I}_\mathrm{\gamma}$ is the intensity of emission in the  PPV cube, obtained from an angle $\gamma$, $\bm{X}$ is the position in the plane of the sky, and $v_\mathrm{los}$ the velocity component along the LOS.
In the particular case of an optically thin medium, with emissivity proportional to the density (e.g. cold \ion{H}{1}) the integrated intensity is proportional to the column density. This is not the case however, when self-absorption is important, or if the emissivity is not linearly proportional to the density.

Since velocity field statistics are the base of most turbulence models, it would be ideal to analyze the statistics on the PPV data to charactrize velocity. However, we cannot measure velocity directly from PPV, so it is common to use velocity centroids instead. We can obtain the velocity centroids of the PPV cubes as the first moment of the spectral lines:
\begin{equation}
    \label{eq:Cent}
        C_\gamma\left(\bm{X} \right)=
        \frac{\int \mathcal{I}_\gamma\left(\bm{X},v_\mathrm{los} \right)\,v_\mathrm{los}\,dv_\mathrm{los}}{I_\gamma\left(\bm{X} \right)}.
\end{equation}

This equation is the usual definition of velocity centroids \citep{1951ZA.....30...17V, 1958RvMP...30.1035M} and we have often called them \emph{normalized} velocity centroids to distinguish them from \emph{unnormalized} velocity centroids, which are not divided by column density but that are easier to treat analytically \citep{2005ApJ...631..320E}.
We should note that, in the case of an optically thin medium, velocity centroids correspond to the mean LOS velocity in the limit of small density fluctuations, however as the density fluctuations and/or self-absorption become important this is no longer the case.

We then calculate the two-point structure-functions of the resulting 2D maps  (e.g. integrated intensity or centroids) for each model, molecular line, velocity mode and orientation ($\gamma$). For instance, for the velocity centroids, the two-point second-order structure-function\footnote{In what follows we will refer to them simply as structure functions.} is defined as:
\begin{equation}
\label{eq:SF}
	\mathrm{SF}_{C,\gamma}(\bm{R}) = \langle [C_\gamma\left(\bm{X}\right)-C_\gamma\left(\bm{X}+\bm{R}\right)]^2\rangle,
\end{equation}
where $\langle\ldots\rangle$ denotes averaging over the entire plane of the sky
\footnote{Turbulence theories consider ensemble averages that are both spatial and temporal. Fortunately, when dealing with scales much smaller than the outer scale the spatial average can be adequate.} ($\bm{X}$).
The lag ($\bm{R}$) is a two-dimensional vector in the plane of the sky. For isotropic turbulence of the structure-function depends on the magnitude of $\bm{R}$ but not on its direction, but in the general case it depends on the direction, and we can exploit this dependence to obtain information about the magnetic field orientation. The latter is possible at scales smaller than the outer scale of turbulence (i.e. the driving scales). At the largest scales, the magnetic field becomes isotropic even at $\mathcal{M}_\mathrm{A,0}=2$ \citep{2020MNRAS.498.1593B} where there is equipartition between kinetic and magnetic energies. At smaller scales the magnetic field dominates and the turbulence becomes anisotropic. A closely related measure is the correlation function $\mathrm{CF}_{C,\gamma}(\bm{R})=\langle C(\bm{X})C(\bm{X+R})\rangle$, which differs from the (second-order) SF basically by a constant \citep[see, for instance][]{2005ApJ...631..320E}.
We must note that computing structure functions in general as in Equation (\ref{eq:SF}) can be computationally intensive as one must perform on the order of $N^2$ operations (where $N$ is the total number of cells). And thus, for large data sets, particularly with 3D simulations it is customary to use a subset, which has to be sufficiently large to converge statistically \citep[see][]{2021NatAs...5..365F}. One way to check the correct convergence of structure functions is to verify the scaling at large separations, where the structure function value tends to two times the variance.
In the present work, however, we use a {\it Fast Fourier Transform} to obtain the correlation function, and from it the structure function. This can only be performed to obtain two-point second-order structure functions, and works best in periodic boundary simulations in a Cartesian grid. We have verified that the structure functions computed this way converge statistically.

Isocontours of structure-function tend to elongate in the direction of the magnetic field component on the plane of the sky (if it is strong enough), or in the case of a very weak magnetic field strength, projected onto the plane of the sky, are circular in shape (isotropic).
That is a graphical way to see whether our models are isotropic or not (figure 2 of \citetalias{2020ApJ...901...11H} has a structure-function example). However, one has to be careful to observe the isocontours at the correct scale, i.e., within the inertial range (in our models we average over separations between $\sim 10$ and $\sim 100$ grid points, which correspond approximately to the inertial range).
We must note that due to the limited resolution, the inertial range in our simulations is not sufficient to resolve the transition between supersonic and subsonic scales \citep[see for instance][]{2021NatAs...5..365F}. Thus the anisotropy that we are measuring is an average of the contribution of the different turbulent cascades. However, such an average should not depend strongly on the detailed transfer across cascades.

To quantify how anisotropic the structure-functions are, we can define an isotropy degree as:
\begin{equation}
    \mathrm{Isotropy\,degree}(\ell) = \frac{\mathrm{SF}(\ell\, \mathbf{\hat{e}}_\|)}{\mathrm{SF}(\ell\, \mathbf{\hat{e}}_\perp)},
\label{eq:ID}
\end{equation}
where $\ell$ is the lag magnitude, which is taken in two directions $\bf{\hat{e}}_\|$ along the direction of the elongation of the contours and $\bf{\hat{e}}_\perp$ perpendicular to them. The isotropy degree will be equal to one if the structure-functions are circular, which means the structure-functions are isotropic, and less than unity if they are anisotropic.

The procedure to obtain the orientation of the plane of the sky component of the magnetic field is straightforward. One should compute the structure function in the area of interest and, if the contours are elongated, the direction of such elongation determines the directions $\bf{\hat{e}}_\|$ and $\bf{\hat{e}}_\perp$. However, a complication present in simulations, due to the limited resolution is that the contours often deviate from the mean magnetic field direction for intermediate to large lags. This departure can be attributed to the lack of inertial range to fully decouple the turbulent motions from the driving mechanism.
A compensation of this effect was addressed in \citet{2018ApJ...865...54Y} and in \citetalias{2020ApJ...901...11H}, involving a rotation of the 2D structure function of the centroids maps, in order to align the iso-contours horizontally (in the direction of the average mean magnetic field projection on the plane of the sky). This correction reduces the error bars in the isotropy degree measured due to scale dependence, but it is difficult to automatize.
In the present work we do not include this correction and we assume that $\bf{\hat{e}}_\|$ corresponds to the horizontal direction ($x$-axis).

%%%%%%%%%%%%%%%%%%%%%%%%%%%%%%%%%%%%%%%%%%%%%%%%%%
\section{MHD Models and Radiative Transfer}\label{sec:models}
\subsection{MHD Models}

We use the grid of numerical simulations of isothermal, compressible, and fully developed MHD turbulence employed in \citetalias{2020ApJ...901...11H}. The simulations were produced with the MHD code used in  \citet{2003MNRAS.345..325C}. The code solves the ideal MHD equations in a periodic Cartesian box of side $2 \pi$, using a second-order-accurate hybrid essentially non-oscillatory (ENO) scheme \citep[see][]{2002PhRvL..88x5001C}. The simulations have a resolution of $512^3$ cells.
%\deleted{The driving is solenoidal, and it is imposed in Fourier space at a fixed wavenumber $k=2.5$ (i.e., to a scale of $1/2.5$ of the computational domain).}

We used the finite-correlated driving scheme described in \citet{yoon2016}. The driving method uses 22 pseudo-random wave vectors in Fourier space in the wavenumber range of $[2,\sqrt{12}]$. In this scheme, the driving correlation time is of the order of the large-eddy turnover time. We must note that the sources of turbulence are diverse (i.e. stellar jets/ouflows, supernova explosions, galactic shear motions, etc), and thus one should expect a mixture of solenoidal and compressive modes \citep[e.g.][]{2020MNRAS.493.4643M}. Moreover, this mixture varies with each particular source of turbulence, evolutionary stage and on scale. In fact, the issue of the role of compressibility can be scale dependent. The ratio of the compressible and solenoidal motions can change as the cascade progresses to smaller scales. The interaction of shocks on large scales with density inhomogeneities can create solenoidal motions at smaller scales. This, however, is impossible to study properly with our existing numerical simulations that have limited inertial range.  For sake of simplicity we restrict our work to a purely solenoidal driving, keeping in mind that it provides a lower limit to the role of compressive modes. The initial magnetic field is uniform and aligned in the $x$-axis. The simulations are evolved for $5$ dynamical times, in which they all reach steady-state \citep[see for instance][]{2021MNRAS.504.4354B}. At that point the rms velocity is also of order unity ($v_\mathrm{rms} \sim 0.7$). Initially, in all simulations the density is uniform with a value $\rho_0 = 1$, the initial gas pressure $P_\mathrm{gas,0}$, and the Alfv\'en speed $v_\mathrm{A,0}=\vert\bm{B}_0\vert/\sqrt{4\pi \rho}$ determine the model. Two important parameters that characterize our simulations are the sonic Mach number and the Alfv\'enic Mach number at the injection scale $\mathcal{M}_\mathrm{s}=  V_\mathrm{L} /c_\mathrm{s}$, and $\mathcal{M}_\mathrm{A,0}= V_\mathrm{L}/v_\mathrm{A,0}$, respectively; where $V_\mathrm{L}=v_\mathrm{rms}$ is the velocity at the injection scale, $c_\mathrm{s}=\sqrt{P/\rho}$ the isothermal sound speed, and, $v_\mathrm{A,0}=\vert\bm{B}_0\vert/\sqrt{4\pi \rho_0}$ the mean field Alfv\'en speed.
%\sout{, and $\left\langle\ldots\right\rangle$ denotes an average over the entire domain.}
%In the  stationary state, the magnitude of the mean magnetic field and its fluctuation could be of the same order.
In the stationary state, the magnitudes of the mean and the fluctuating magnetic fields are of the same order when $\mathcal{M}_\mathrm{A,0} \sim 1$; while the magnitudes of the fluctuating magnetic fields are smaller than those of the mean fields when $\mathcal{M}_\mathrm{A,0}<1$.
However, the mean magnetic field remains aligned in the original orientation (along $x$-axis).

\begin{deluxetable*}{lCCCCCCCC}
\tablecaption{Grid of MHD simulations and average optical depths}
\label{tab:models}
\tablewidth{0pt}
\tablehead{
    \colhead{Model}
  & \colhead{$v_{\mathrm{rms}}$}
  & \colhead{$\mathcal{M}_\mathrm{A,0}$}
  & \colhead{$\mathcal{M}_\mathrm{s}$}
  & \colhead{$\beta$}
  & \colhead{$\langle\tau_\mathrm{^{12}CO}\rangle$\tablenotemark{*}}
  & \colhead{$\langle\tau_\mathrm{^{13}CO}\rangle$\tablenotemark{*}}
  & \colhead{$\langle\tau_\mathrm{C^{18}O}\rangle$\tablenotemark{*}}
  & \colhead{$\langle\tau_\mathrm{C^{17}O}\rangle$\tablenotemark{*}}
}
\startdata
M1  & \sim 0.79 & \sim 7.90  & \sim 7.90  & 2      & 38.76~(44.41) & 7.776~(9.007) & 0.077~(0.089) & 0.014~(0.016)\\
M2  & \sim 0.78 & \sim 7.80  & \sim 2.47  & 20     & 39.93~(46.13) & 8.340~(9.767) & 0.079~(0.092) & 0.015~(0.017)\\
M3  & \sim 0.77 & \sim 7.71  & \sim 0.77  & 200    & 40.69~(48.00) & 8.098~(9.539) & 0.081~(0.096) & 0.015~(0.018)\\
M4  & \sim 0.75 & \sim 7.49  & \sim 0.53  & 400    & 41.85~(49.26) & 8.168~(9.678) & 0.084~(0.098) & 0.015~(0.018)\\
\hline
M5  & \sim 0.72 & \sim 1.43  & \sim 7.16  & 0.08   & 42.94~(50.21) & 9.60~(11.248) & 0.086~(0.100) & 0.016~(0.018)\\
M6  & \sim 0.69 & \sim 1.37  & \sim 2.17  & 0.8    & 44.08~(50.97) & 10.1~(11.790) & 0.088~(0.102) & 0.016~(0.019)\\
M7 	& \sim 0.67 & \sim 1.34  & \sim 0.67  & 8      & 43.74~(51.09) & 10.1~(11.851) & 0.088~(0.102) & 0.016~(0.019)\\
M8	& \sim 0.66 & \sim 1.32  & \sim 0.47  & 16     & 43.98~(51.68) & 10.1~(11.823) & 0.088~(0.103) & 0.016~(0.019)\\
\hline
M9  & \sim 0.75 & \sim 0.75  & \sim 7.54  & 0.02   & 41.09~(48.33) & 9.88~(11.599) & 0.082~(0.097) & 0.015~(0.018)\\
M10 & \sim 0.72 & \sim 0.72  & \sim 2.28  & 0.2    & 41.58~(48.32) & 9.85~(11.680) & 0.083~(0.096) & 0.015~(0.018)\\
M11 & \sim 0.76 & \sim 0.76  & \sim 0.76  & 2      & 39.05~(44.82) & 9.69~(11.415) & 0.078~(0.089) & 0.014~(0.016)\\
M12 & \sim 0.77 & \sim 0.77  & \sim 0.54  & 4      & 37.69~(44.50) & 9.76~(11.541) & 0.076~(0.089) & 0.014~(0.016)\\
\hline
M13 & \sim 0.76 & \sim0.38   & \sim 7.62  & 0.005  & 40.81~(46.66) & 10.4~(12.049) & 0.081~(0.093) & 0.015~(0.017)\\
M14 & \sim 0.77 & \sim0.39   & \sim 2.45  & 0.05   & 43.33~(49.30) & 11.3~(13.174) & 0.086~(0.098) & 0.016~(0.018)\\
M15 & \sim 0.84 & \sim0.42   & \sim 0.84  & 0.5    & 40.75~(47.25) & 9.73~(11.387) & 0.081~(0.094) & 0.015~(0.017)\\
M16 & \sim 0.83 & \sim0.42   & \sim 0.59  & 1      & 40.85~(47.12) & 9.44~(10.882) & 0.082~(0.094) & 0.015~(0.017)\\
\hline
M17 & \sim 0.80 & \sim 0.27  & \sim 8.05  & 0.0022 & 41.75~(47.82) & 9.85 (11.756) & 0.084~(0.095) & 0.015~(0.018)\\
M18 & \sim 0.81 & \sim 0.27  & \sim 2.57  & 0.022  & 42.87~(49.04) & 10.6~(12.670) & 0.086~(0.098) & 0.016~(0.018)\\
M19 & \sim 0.84 & \sim 0.28  & \sim 0.84  & 0.22   & 40.11~(46.80) & 9.06~(10.569) & 0.080~(0.093) & 0.015~(0.017)\\
M20 & \sim 0.82 & \sim 0.27  & \sim 0.58  & 0.44   & 40.06~(47.03) & 8.560~(9.981) & 0.081~(0.094) & 0.015~(0.017)\\
\hline
M21 & \sim 0.86 & \sim 0.17  & \sim 8.61  & 0.0008 & 42.12~(48.47) & 9.61~(11.383) & 0.084~(0.097) & 0.016~(0.018)\\
M22 & \sim 0.85 & \sim 0.17  & \sim 2.70  & 0.008  & 41.22~(47.02) & 8.701~(9.976) & 0.082~(0.094) & 0.015~(0.017)\\
M23 & \sim 0.83 & \sim 0.17  & \sim 0.83  & 0.08   & 40.85~(47.16) & 8.55~(10.029) & 0.081~(0.094) & 0.015~(0.017)\\
M24 & \sim 0.81 & \sim 0.16  & \sim 0.57  & 0.16   & 43.64~(50.63) & 8.048~(9.429) & 0.087~(0.101) & 0.016~(0.019)
\enddata
\tablecomments{$\mathcal{M}_\mathrm{A,0}= v_\mathrm{rms}/v_\mathrm{A,0}$, $\mathcal{M}_\mathrm{s}= v_\mathrm{rms}/c_\mathrm{s}$, and $\beta = P_\mathrm{gas}/P_\mathrm{mag} =2 \left( \mathcal{M}_\mathrm{A,0}^2/ \mathcal{M}_\mathrm{s}^2\right) $. All the models have a resolution of $512^3$ cells.}
\tablenotetext{*}{The first value is the optical depth averaged in velocity  between $v_0\pm \sigma_\mathrm{v}$ over the entire plane of the sky, and the value inside parenthesis is the optical depth at $v=v_0$, also averaged over the plane of the sky ($v_0$ is the mean line of sight velocity and $\sigma_\mathrm{v}$ the turbulent velocity dispersion).}
\end{deluxetable*}

The models are summarized in Table \ref{tab:models}, where model name, the resulting Mach numbers, and the plasma $\beta=P_\mathrm{gas}/P_\mathrm{mag}=2(\mathcal{M}_\mathrm{A,0}^2/\mathcal{M}_\mathrm{s}^2)$ are listed in the first four columns. The Mach numbers cover a wide range of subsonic and supersonic, along with sub-Alfv\'enic and super-Alfv\'enic turbulence regimes, and the plasma $\beta$ includes regimes in which the magnetic fields are dynamically dominant ($\beta \ll 1$) or dynamically unimportant ($\beta \gg 1$).  In the table we also list a series of optical depths which will be discussed in a later section.%later in a Section \ref{sec:radiative}.

%Also, the $\mathrm{CO}$ isotopologues optical depths, $\tau$, are listed.

\subsection{Separation of MHD modes}\label{sec:modes}

As done in \citetalias{2020ApJ...901...11H} for the optically thin case, we want to study the anisotropy contribution of each of the different MHD modes (Alfv\'en, slow and fast modes). We follow the procedure described in \citet{2002PhRvL..88x5001C,2003MNRAS.345..325C} to separate the original velocity field in each of the MHD modes. In a nutshell, the separation consists in taking the Fourier components of the velocity; projecting them in the directions of the displacement vector of the Alfv\'en, slow and fast modes \citep[see figure 1 and appendix A in][for further detail]{2003MNRAS.345..325C}. Finally, we transform back to space the Fourier projection of each mode, and we obtain the three components ($x$, $y$, $z$) of a new velocity field associated to each mode (Alfv\'en, slow, and fast). With these velocity fields (three per model in addition to the original velocity field) we construct PPVs for different molecular lines and orientations,  and from them obtain 2D maps of integrated intensities and velocity centroids.

\subsection{Radiative Transfer}
\label{sec:radiative}

We take the results of the simulations: density, temperature, and velocity field (either the original or one that corresponds to a particular MHD mode) and pass them to a radiation transfer code in order to obtain synthetic observations (PPV cubes) in various CO molecular lines.
This step is made in post-processing, and it is based in the {\it Python Radiative Transfer Emission code} \citep[\textsc{pyrate},][]{2018MNRAS.478.2056T}.

%We post-process the Cartesian cubes using the Python Radiative Transfer Emission code \citep[\text sc{pyrate},][]{2018MNRAS.478.2056T}. The \textsc{pyrate} code is a multi-level, non-local thermodynamic equilibrium (non-LTE) line radiative transfer code.
Due to the slow integration of the radiative transfer equation in the \textsc{pyrate} code, we wrote a \textsc{fortran} version suited for our particular needs.
We generate four synthetic emission lines of $\mathrm{CO}$ isotopologues, i.e., $\mathrm{^{12}CO}$,    $\mathrm{^{13}CO}$, $\mathrm{C^{18}O}$, and $\mathrm{C^{17}O}$.

The radiative transfer code takes the density, temperature (constant in our case), velocity field and molecular abundance as input values. Molecular data are initially loaded into the code, including the central emission frequency of each spectral line, the Einstein coefficients as well as collisional excitation/de-excitation coefficients.

\subsection{Molecular Initial conditions}

We scaled our simulations to a cubic Cartesian box of $L=5$ pc on each side, mean numerical density of $n = 275$ cm$^{-3}$ (where we assume that $98\%$ are hydrogen nuclei particles and $\mu = 2.36\,m_\mathrm{H}$.
%, see \citealt{2002A&A...391..295O}).
The gas temperature is taken to be $T=10$ K (typical in molecular clouds), the spectral resolution is $0.012$ km s$^{-1}$, and the sound speed of $c_s=0.2$ km s$^{-1}$.
%However, throughout the code, we use cgs units. \\

The fractional abundances of the $\mathrm{CO}$ isotopologues $\mathrm{^{12}CO}$, $\mathrm{^{13}CO}$, $\mathrm{C^{18}O}$, and $\mathrm{C^{17}O}$ relative to H$_2$ are $1.7\times 10^{-4}$, $2\times 10^{-5}$, $3.7\times 10^{-7}$, and $6.7\times 10^{-8}$, respectively.
We choose the lowest-transition $\mathcal{J} = 1\xrightarrow{}0$ of the $\mathrm{CO}$ isotopologues \citep{estalella_anglada_2008}.

\subsection{Radiative Transfer equation}

The code uses the non-relativistic, time-independent radiation transfer equation. The numerical integration procedure of the transport equation is the same as in \citet{1986ASIC..188..141Y}, where the line and continuous emission contributions are taken separately. Integration of said equation from point $a$ to point $b$ yields
\begin{equation}
        I_b = \frac{\left(e^{-\uptau_b^C}-p\right)I_a + pS_a^L +qS_b^L + S^k}{1+q}
        \label{eq:RT}
\end{equation}
where the $a$ and $b$ subscripts refer to quantities measured at such positions, $L$ and $C$ superscripts denote the contributions of line and dust continuum emission, respectively; $I$ is the radiative intensity, $S^L$ the source function for the line emission, and $\uptau^C$ is the optical depth for continuum emission. The other quantities are given by
\begin{equation}
    q = \frac{\uptau^L_{b}}{1+e^{-\uptau^L_{b}}},
    \label{eq:q}
\end{equation}
\begin{equation}
    p = q(e^{-\uptau^L_{b}-\uptau^C_{b}}),
    \label{eq:p}
\end{equation}
\begin{equation}
    S^k = e^{-\uptau^C_{b}}\int_{a}^{b}\kappa^CS^C\exp\left(\int_{a}^s\kappa^C\mathrm{d}s'\right)\mathrm{d}s,
    \label{eq:Sk}
\end{equation}
\begin{equation}
    \uptau_b^L = \int_a^b\kappa^L\mathrm{d}s.
    \label{eq:tauL}
\end{equation}

Here, $S^C$ is the source function for dust continuum emission, and $\uptau^L$ is the optical depth of the line. The length along the LOS is denoted as $s$, $\kappa^C$ is the extinction coefficient for continuum emission that we can obtain from the dust model based in \citet{1993A&A...279..577P} which is appropriate for dense molecular cloud conditions.

The model includes three components of dust grains, namely amorphous carbon grains, silicate grains, and dirty ice (i.e., water, ammonia, and carbon particles) coated silicate grains.
The number of grains per gram of gas are $\log N_d = 14.17$, $1.0$, and $12.001$ g$^{-1}$ for the amorphous carbon, silicate, and dirty ice-coated silicate grains, respectively. The sublimation temperature of each type of grain is $~2000$, $1500$, and $125$ K, respectively. The extinction coefficient for dust emission is then computed as:

\begin{equation}
    \kappa^C = \sum_{d=1}^3n_dQ_d^{ext}\pi r^2_d,
    \label{eq:kC}
\end{equation}

where $n_d$ is the number density of each grain component, $Q_d^{ext}$ is the efficiency factor for extinction for each grain component and $r_d$ the radius of the (spherical) grains, which is taken to be $\sim 10$ nm for the carbon, $\sim 50$ nm for the silicate, and $\sim 60$ nm for the dirty ice-coated silicate grains. Finally, $\kappa^L$ the line extinction coefficient is computed as:

\begin{equation}
    \kappa^L = n_mB_{mn}\frac{h\nu_0}{4\pi}\left[1-\frac{n_ng_m}{n_mg_n}\right]\varphi(\nu),
    \label{eq:kL}
\end{equation}

where $n$ is the population density of the upper and lower level (denoted with the subscripts $n$ and $m$, respectively). $B_{nm}$ is the Einstein $B$ coefficient, $h$ the Planck constant, $\nu_0$ the rest frequency of the line, and $\varphi(\nu)$ is the normalized profile function (its integral over frequency is unity).

 The resulting emission is obtained proceeding along the LOS considering pairs of cells advancing from the back of the ``cloud'' towards the observer (for more details see  \citealt{1986ASIC..188..141Y}).
%As a result, we can obtain $I_b$ by calculating Eqs. (\ref{eq:q}) through (\ref{eq:kL}) and put them in Eq. (\ref{eq:RT}) and perform the ray tracing \\

%%%%%%%%%%%%%%%%%%%%%%%%%%%%%%%%%%%%%%%%%%%%%%%%%%%%%
%%%%%%%%%%%%%%%%%%%%%%%%%%%%%%%%%%%%%%%%%%%%%%%%%%%%%

\section{Results}\label{sec:results}

\subsection{Sample spectral lines and baseline correction}

After the radiation transfer post-processing step, we are left with a series of PPV data cubes with spectral lines that range from nearly Gaussian and optically thin for $\mathrm{C^{17}O}$ to saturated, optically thick for $\mathrm{^{12}CO}$.
The simulated PPV data were obtained with a fixed number of (100) velocity channels,  using a velocity range from $- 0.6~\mathrm{km\,s^{-1}}$ to $0.6~\mathrm{km\,s^{-1}}$  so that we capture the whole emitting material in all the models (this corresponds to a uniform velocity resolution of $\Delta v= 0.012~\mathrm{km\,s^{-1}}$).

\begin{figure}[htp]
    \centering
    \includegraphics[scale=0.55]{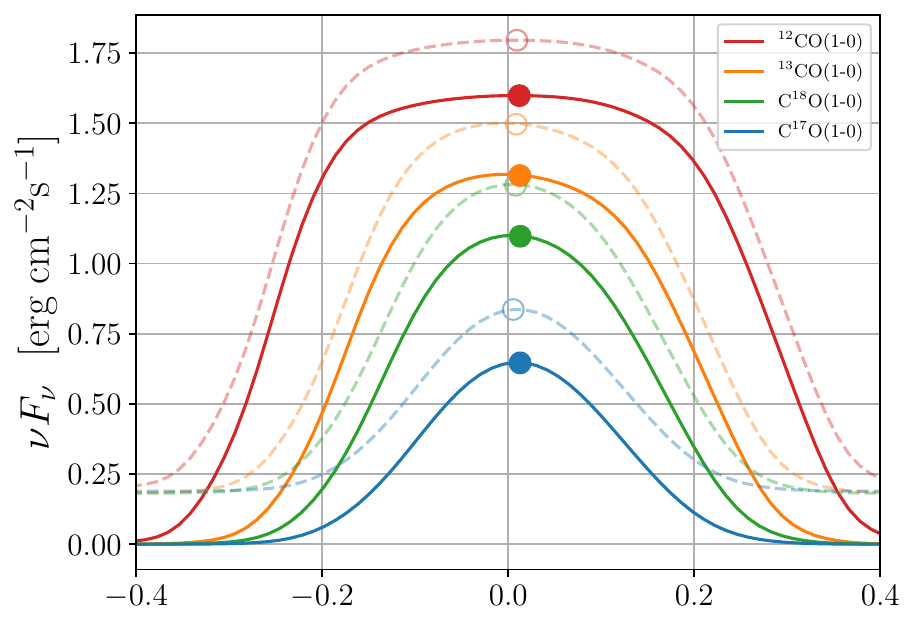}
	\includegraphics[scale=0.55]{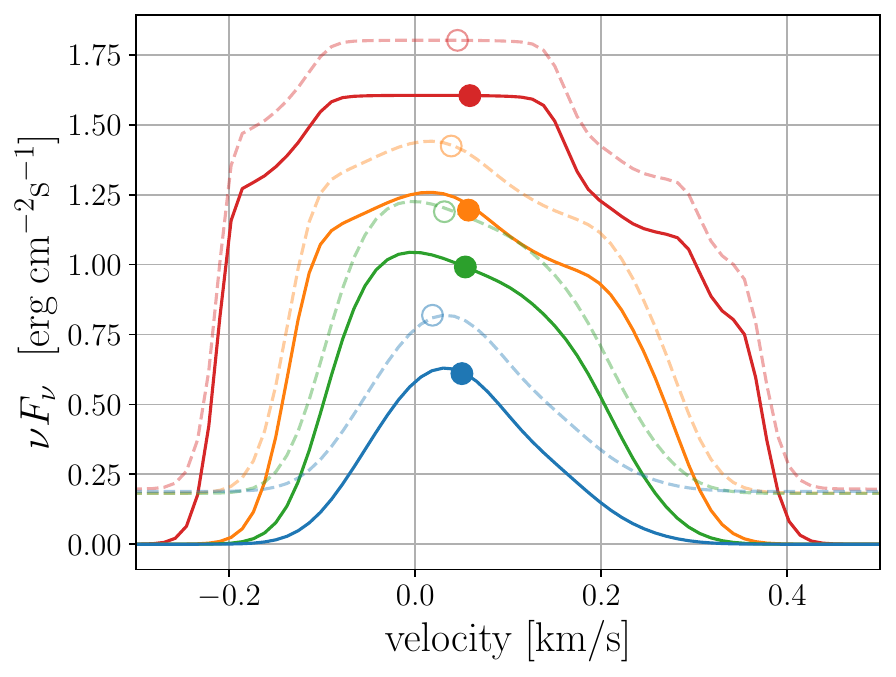}
	\caption{Spectral Lines of model M13 ($\mathcal{M}_\mathrm{A,0}\sim 0.4$ and $\mathcal{M}_\mathrm{s}\sim 7.6$) to the $\mathrm{^{12}CO}$ isotopologues namely, $\mathrm{^{12}CO}$ (red line), $\mathrm{^{13}CO}$ (orange line), $\mathrm{C^{18}O}$ (green line), and $\mathrm{C^{17}O}$ (blue line). The top plot shows the plane of the sky average spectral line, and the bottom panel shows spectral lines along a position in the sky which has a large velocity. In both panels, the solid lines represent spectra with baseline correction while the dashed lines are the spectral lines before such correction. The circles show the velocity centroids position.}
    \label{fig:SLines}
\end{figure}

In Figure \ref{fig:SLines}, we show examples of the spectral lines for the different molecules for a strongly magnetized and supersonic model, M13. In the top panel we show the average (over the plane of the sky) spectral line and in the bottom panel we present the spectral line towards an arbitrary position with high centroid velocity.% of high velocity.
One can see clearly that the emission from $\mathrm{^{12}CO}$ is saturated while the other lines become progressively optically thin for the rest of the molecules, with $\mathrm{C^{17}O}$ being the thinnest.

Since the turbulence is more or less homogeneous the average spectral lines (top panel of Figure \ref{fig:SLines}), are symmetrical around the mean velocity of zero. However this is not true for any given line, e.g. the bottom panel.

The resulting spectral lines are shown with dashed lines, and the position of the centroids with open symbols.
While the centroids for the average spectral lines coincide rather well with the mean velocity it is clear how the centroids are affected by the continuum in an arbitrary line of sight (exemplified in the bottom panel).

This \emph{baseline correction}, a pre-processing observational technique, to separate the valid spectroscopic signal from interference effects or remove background effects or noise is standard \citep{LILAND201151} when dealing with real data. However, it is sometimes overlooked with simulated data. This example shows the need for properly taking it into account with synthetic data as well, which we did for all the models. The results after the baseline correction are included in Figure \ref{fig:SLines}, depicted by solid lines and filled circles.

\subsection{Integrated intensity}

\begin{figure*}[htp]
    \centering
	\includegraphics[scale=0.95]{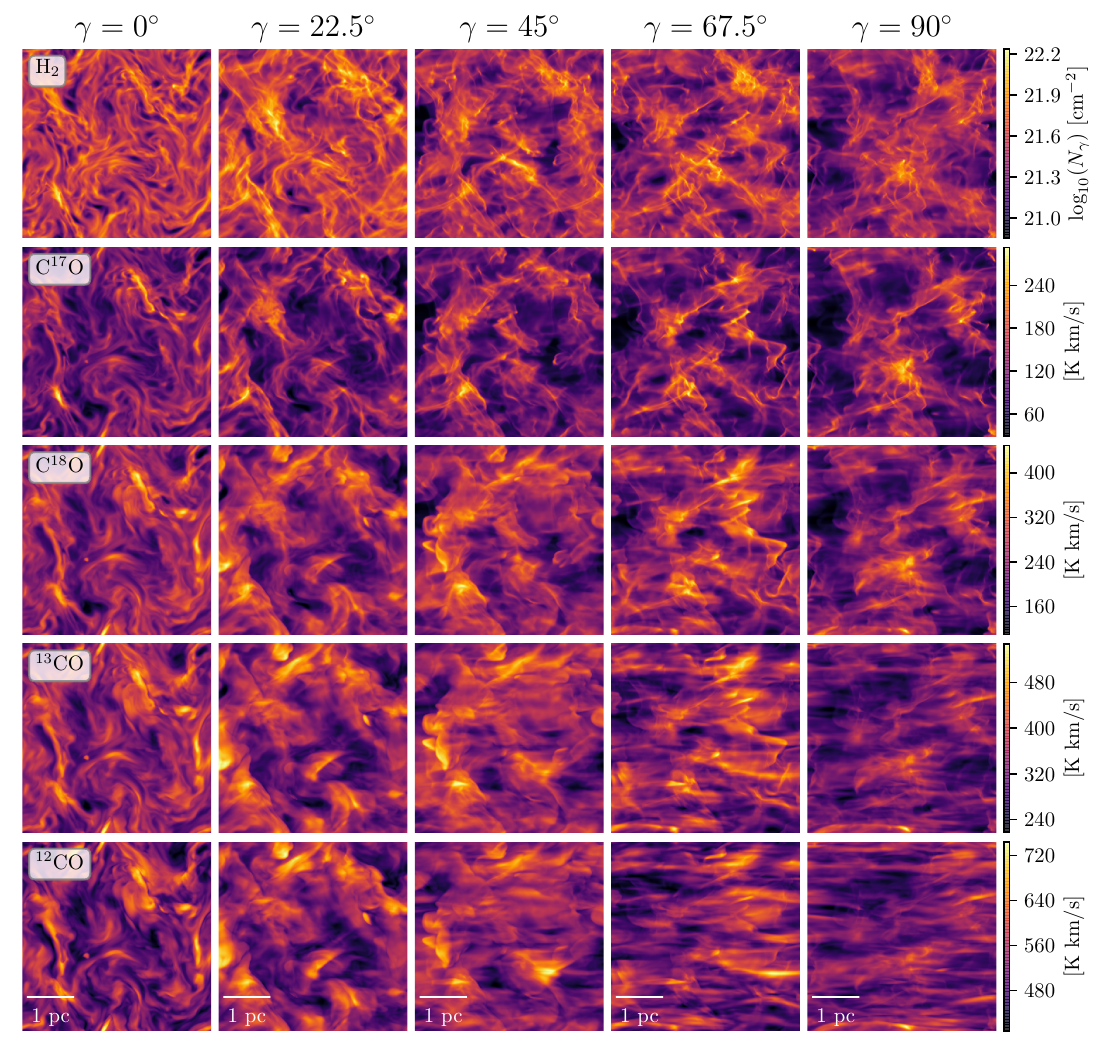}
	\caption{Integrated intensity maps obtained from model M13 ($\mathcal{M}_\mathrm{A,0}\sim 0.4$ and $\mathcal{M}_\mathrm{s}\sim 7.6$), with dimensions $\mathrm{L=5~pc}$ per side, mean numerical density $n=275~\mathrm{cm}^{-3}$ (mass $M \sim 2000~\mathrm{M}_\odot$) and temperature $T = 10~\mathrm{K}$. By rows, from top to bottom (also in increasing optical thickness): column density of $\mathrm{H}_2$ molecule and  $\mathrm{C^{17}O}$, $\mathrm{C^{18}O}$, $\mathrm{^{13}CO}$, and $\mathrm{^{12}CO}$ isotopologues, as shown by the white legend in the upper-left corner. By columns, from left to right, the maps obtained with $\gamma=0\degr,~22.5\degr~45\degr,~67.5\degr,~90\degr$, respectively, where $\gamma$ is the angle between LOS and the mean magnetic field, $\mathrm{B}\mathbf{\hat{x}}$.}
    \label{fig:CD}
\end{figure*}

In Figure \ref{fig:CD}, we show the integrated intensity (which is proportional to column density in the optically thin limit) for model M13 ($\mathcal{M}_A \sim 0.38$ and $\mathcal{M}_s \sim 7.62$). Each column shows the different LOS where $\gamma$ is the angle between the LOS and the direction of the mean magnetic field projected onto the plane of the sky, so from left to right, we have maps in which the LOS changes from parallel to the magnetic field ($\gamma = 0\degr$)  to perpendicular to it ($\gamma = 90\degr$). By rows we have the $\mathrm{H}_2$ column density (top row) and the integrated emission from the $\mathrm{CO}$ isotopologues $\mathrm{C^{17}O}$, $\mathrm{C^{18}O}$, $\mathrm{^{13}CO}$, and $\mathrm{^{12}CO}$ in downward progression, respectively.

Let us turn our attention to the first row of Figure \ref{fig:CD}, which contains the column density of $\mathrm{H}_2$ calculated assuming that it is optically thin (no radiation transfer or self-absorption included). The structures in the 2D projection show differences with the viewing angle (indicated at the top of each column) that changes from being parallel to the mean magnetic field in the leftmost panel to perpendicular to it in the rightmost panel. Although the structures have a different morphology that can be identified visually, there is not an obvious alignment with the mean magnetic field for the $\mathrm{H}_2$  maps. The same tendency was reported in \citetalias{2020ApJ...901...11H}.

%A very similar behavior is found for the optically thin $\mathrm{C^{17}O}$ and $\mathrm{C^{18}O}$ lines, especially for small viewing angles ($\gamma \leq 45\degr$). For larger viewing angles there are subtle differences in the morphology, and some indications of alignment with the mean magnetic field start to appear.

%At the same time for larger values of $\gamma$ the integrated intensity maps become clumpier, and in some cases they start to show an indication of alignment with the mean magnetic field (a few horizontal filament-like structures), the same behavior was found \citetalias{2020ApJ...901...11H}.

%From this figure, we can see focused in the $\mathrm{H}_2$ molecule (fist row). The structures of the intensity integrated maps parallel to the magnetic field (first column, $\gamma=0\degr$) do not have a preferential direction. In a way, the next map to right ($\gamma=22.5\degr$) has the same behavior than the first map. The difference is observed from the third to the fifth columns ($\gamma=45\degr$, $\gamma=67.5\degr$, and $\gamma=90\degr$, respectively). The structures begin to aligned in the direction of the projected mean magnetic field in the plane of the sky (horizontally).  We observed the same behavior in \citet{2020ApJ...901...11H}.

Now, if we compare the top row ($\mathrm{H}_2$) with the rest (from top to bottom, increasing in optical thickness) and when $\gamma\geq 45\degr$ (in columns), we see that, for the $\mathrm{C^{17}O}$ and $\mathrm{C^{18}O}$ isotopologues the morphology of the emitting structures remains quite similar.% cases (and remarkably similar to that of $\mathrm{H}_2$).
However, in the last two rows ($\mathrm{^{13}CO}$ and $\mathrm{^{12}CO}$) we observe a very clear alignment of structures along the mean magnetic field direction (horizontally aligned) for large viewing angles.

The reason for this change is that, according to the \citetalias{1995ApJ...438..763G} model, small-scale motions are more aligned with the magnetic field as the kinetic to magnetic energy ratio is higher compared to larger scales. In optically thick tracers, in which one can only probe up to a limited depth (thus largest scales are not sampled), the resulting maps are expected to reflect the higher alignment associated with small scales.

\subsection{Original Velocity Centroids}

As we saw in Section \ref{sec:centroids}, we can obtain the velocity centroids from the PPV data using Eq. \ref{eq:Cent}, as we do with observations.

\begin{figure*}[htp]
    \centering
	\includegraphics[scale=0.92]{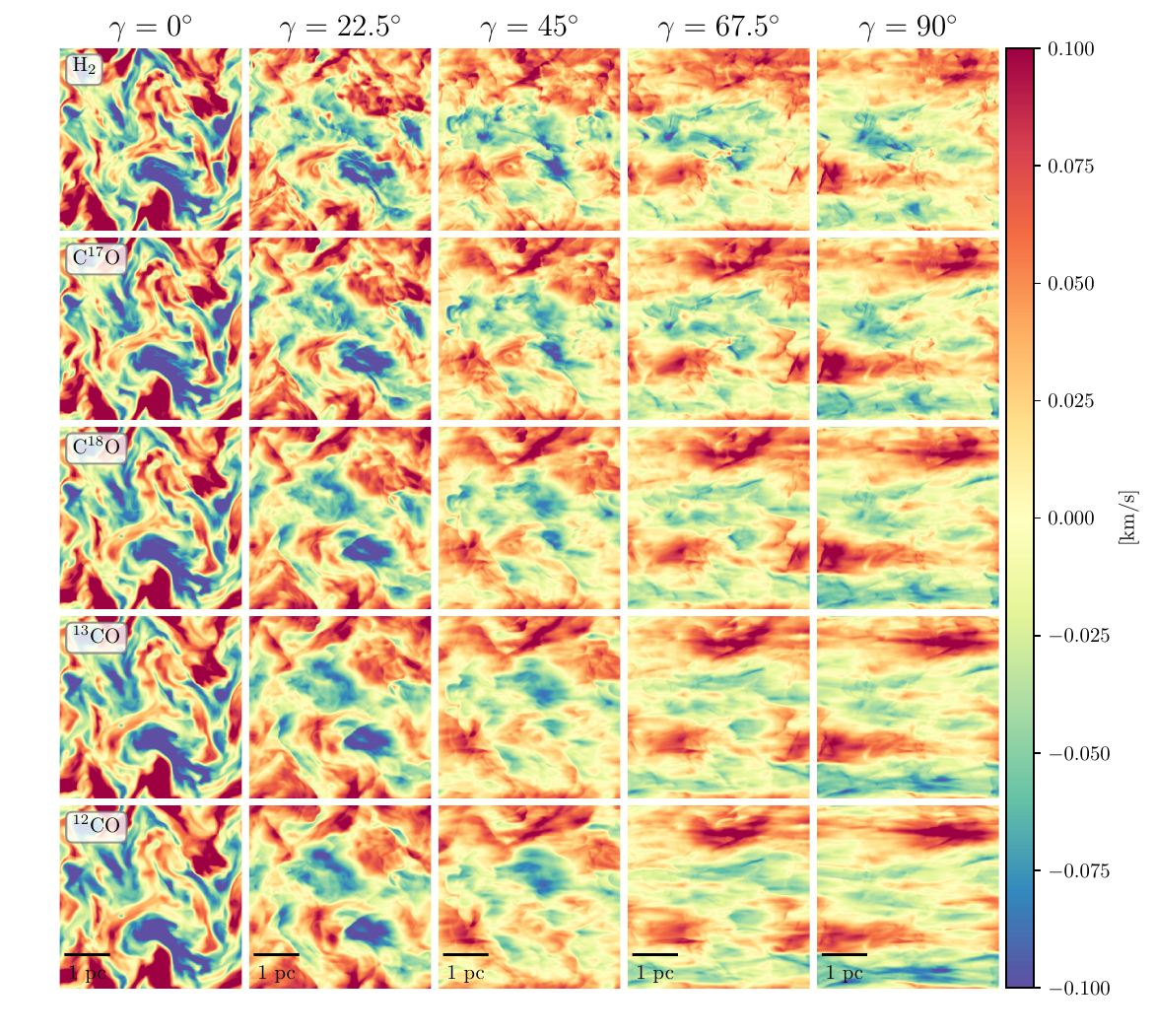}
	\caption{Velocity centroids 2D maps from model M13 ($\mathcal{M}_\mathrm{A,0}\sim 0.4$ and $\mathcal{M}_\mathrm{s}\sim 7.6$). The plot has the same arrangement of Figure \ref{fig:CD}. The color-bar indicate the velocity in $\mathrm{km~s^{-1}}$.}
    \label{fig:OVC}
\end{figure*}

Figure \ref{fig:OVC} shows M13 velocity centroids with the different tracers by rows, and $\gamma$ angles by columns (as in Figure $\ref{fig:CD}$). The color-bar is the same in all the plots, showing velocities between $-0.1$ and $0.1$ km s$^{-1}$.

First let us turn our attention to the first row that corresponds to the $\mathrm{H_2}$ molecule. We can see that the parallel LOS to the magnetic field the velocity centroids do not align in either direction.
However, for $\gamma\geq45\degr$, we see that the velocity centroids do show structures aligned with the magnetic field direction. This trend was also seen in \citetalias{2020ApJ...901...11H}.

On the following four rows, which correspond to the $\mathrm{CO}$ isotopologues, from top to bottom in increasing optical thickness. Similarly to the $\mathrm{H_2}$ molecule, we do not see a preferential direction if the LOS is parallel to the magnetic field. However, for  $\gamma\geq 45\degr$, we observe that several structures in the velocity centroids maps align in the horizontal direction.
Moreover, in the optically thick tracers (the fourth and the fifth rows), the alignment becomes more obvious.

In general we see that the anisotropy in velocity centroids is less restricted to large viewing angles compared to integrated intensity maps. At the same time they are also less affected by radiation transfer effects. This will be quantified in Section \ref{sec:ID}.

\subsection{MHD-modes Velocity Centroids}

We present in this section results with the velocity cubes of the different MHD modes, as mentioned before we calculate the velocity centroids for all models and all LOS (viewing angles).

\begin{figure*}[htp]
    \centering
	\includegraphics[scale=1.1]{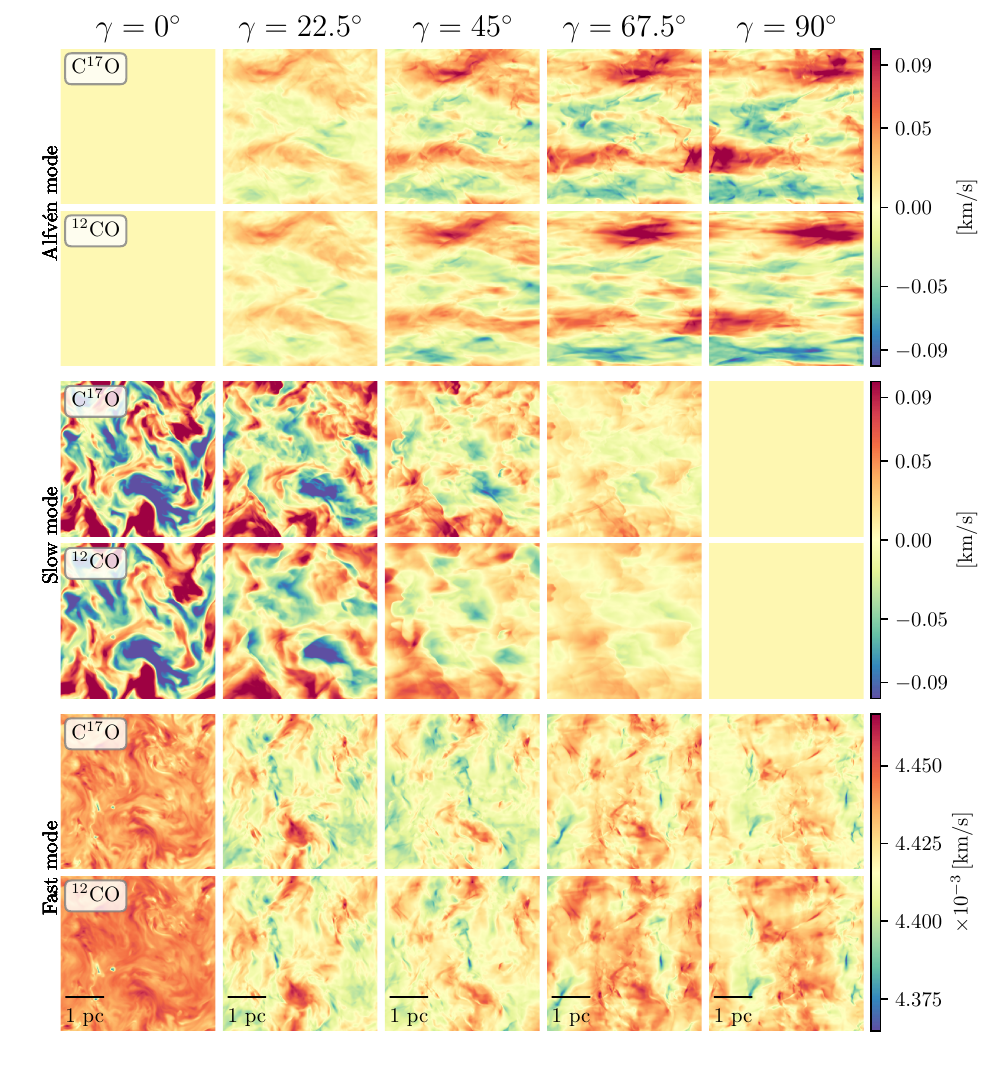}
	\caption{Velocity centroids of MHD Alfv\'en (rows 1 and 2), Slow (rows 3 and 4), and Fast (rows 5 and 6) modes of the model M13 ($\mathcal{M}_\mathrm{A,0}\sim 0.4$ and $\mathcal{M}_\mathrm{s}\sim 7.6$). From top to bottom, the odd rows are optically thin ($\mathrm{C^{17}O}$ isotopologues) and the even rows are optically thick ($\mathrm{^{12}CO}$ isotopologues). The columns are the LOS, like Figure \ref{fig:OVC}. The color-bar indicates the velocity in km s$^{-1}$ units.}
    \label{fig:MHDmodes}
\end{figure*}

In Figure \ref{fig:MHDmodes}, we show the velocity centroids for the different the MHD modes, Alfv\'en (first and second rows), slow (third and fourth rows), and fast modes (fifth and sixth rows). The odd rows correspond to the optically thin cases ($\mathrm{C^{17}O}$), and the even rows are the optically thick cases ($\mathrm{^{12}CO}$). As in the two previous figures, the columns correspond to the different LOS (from parallel to perpendicular magnetic field, as labeled at the top of each column).

Comparing the maps in Fig. \ref{fig:MHDmodes} with their counterparts obtained with the original velocity cubes in Fig. \ref{fig:OVC}, we can see that the Alfv\'en mode dominates the contribution of the centroids when the LOS is perpendicular to the mean magnetic field (since the Alfv\'en waves are transverse), while a significant contribution of the slow modes occurs when the LOS is parallel to the mean-field (thus are isotropic). The fast modes have a smaller velocity-range than the other modes (see the scale in the color-bar), and they do not contribute significantly. In the Alfv\'en modes, we see the same trends of the original velocity centroids, that is, an alignment for angles $\gamma \gtrsim 45\degr$ which is more pronounced for optically thick tracers.
At the same time, the amplitude of Alfv\'en mode increase with $\gamma$, while for the slow mode decrease with $\gamma$ (as previously found in \citetalias{2020ApJ...901...11H} for optically thin media).

\subsection{Average Isotropy degree}\label{sec:ID}

Once we have the  2D maps (integrated intensity and velocity centroids), we calculate their structure-function and from these the average isotropy of each model as described (for more details see \citetalias{2020ApJ...901...11H}, and figures 2 and 3 therein).

%\begin{figure*}[htp]
%    \centering
%	\includegraphics[scale=1]{fig5.pdf}
%	\caption{Average Isotropy Degree of integrated Intensity vs. Alfv\'enic Mach number for all models. In rows, from top to bottom, we show the different tracers, from optically thin to thick (as previous figures). In columns from left to right we have different LOS orientations, with viewing angles $\gamma$ from $0\degr$ to $25.5\degr$, $45\degr$, $67.5\degr$, and $90\degr$. Each symbol represents a model (see Table \ref{tab:models}). The shape of the symbols group models with low $\beta$ (magnetically dominated as circles), intermediate $\beta$ (squares), and high $\beta$ (mostly hydrodynamical, diamonds). The color of the symbols represents the sonic Mach number, $\mathcal{M}_s$, as labeled in the legend in the top-left panel. \dvd{The error bars represent the maximum and minimum values of isotropy degree}}
%    \label{fig:AidI}
%\end{figure*}

\begin{figure*}[htp]
    \centering
	\includegraphics[scale=1]{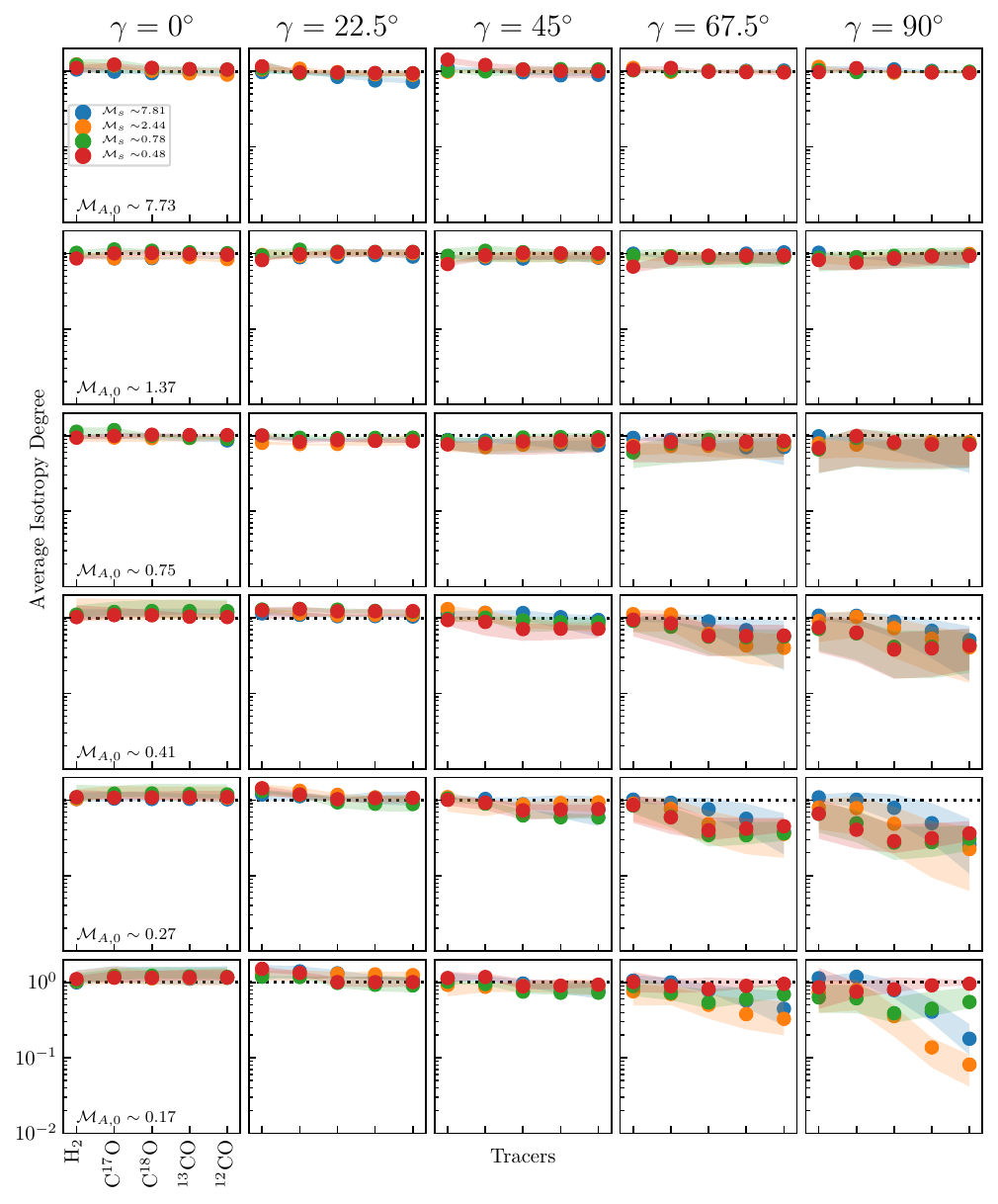}
	\caption{Average Isotropy Degree of Integrated Intensity for the different tracers in all models, from optically thin ($\mathrm{H_2}$) to thick ($\mathrm{^{12}CO}$). In rows, from top to bottom, the Alfv\'enic Mach number are shown in descending order. From left to right, we display different LOS orientations in columns, with viewing angles $\gamma$. The colors indicate the sonic Mach number, as shown in the top-left panel. The shadows represent the maximum and minimum isotropy degree values for each model.}
    \label{fig:AidI}
\end{figure*}

To compare across different models and viewing angles we compute for each case the isotropy degree from the structure-function of the maps as defined in Eq. \ref{eq:ID} and average over the inertial range (from $10$ cells to $1/5$th of the computational domain).

The average isotropy degree of the integrated intensity in all the models is shown in Figure \ref{fig:AidI}. Each symbol therein corresponds to a model and a viewing angle; the results are plotted as a function of the tracers, i.e. the $\mathrm{H_2}$ molecule and the different $\mathrm{CO}$ isotopologues from optically thin to thick.
Each row, from top to bottom, corresponds to a similar Alfv\'enic Mach number, $\mathcal{M}_A$ in descending order (indicated in the label at the bottom left corner in the first panel of each row) . In columns, from left to right, we explore different viewing angles, from LOS parallel to $\mathbf{B}_0$ ($\gamma=0\degr$) to perpendicular to $\mathbf{B}_0$ ($\gamma=90\degr$). The models are also distinguishable by their sonic Mach number, $\mathcal{M}_s$ (in colors, see legend in the upper-left plot). The dotted horizontal line indicates the isotropy degree equal to unity (i.e. when the iso-contours from the structure-functions are circular and therefore considered isotropic).

In Figure \ref{fig:AidI}, we can see that, regardless of the tracer, for small viewing angles ($\gamma \leq 22.5\degr$) the structure-functions are isotropic. For $\gamma = 45\degr$ we start to notice some departure of isotropy (i.e. anisotropy, which becomes more apparent for  $\gamma \geq 67.5\degr$), mainly for those tracers that are optically thick and a $\mathcal{M}_{A,0}$ less than unity. In addition, the more magnetized models show increasingly anisotropy with the viewing angle $\gamma$.
Also, as it could be anticipated by the filamentary structures noticed in Figure \ref{fig:CD} the anisotropy increases with the optical thickness of the tracer for larger viewing angles ($\gamma \geq 67.5^{\circ}$). Interestingly, there is a noticeable difference between the super- and sub-sonic models in this regard: in the former case (blue and orange symbols) the anisotropy keeps increasing with opacity, whereas in the former (red and orange symbols) the anisotropy seems to saturate for $\mathrm{C^{18}O}$, $\mathrm{^{13}CO}$) and $\mathrm{^{12}CO}$).

\begin{figure*}[htp]
    \centering
	\includegraphics[scale=1]{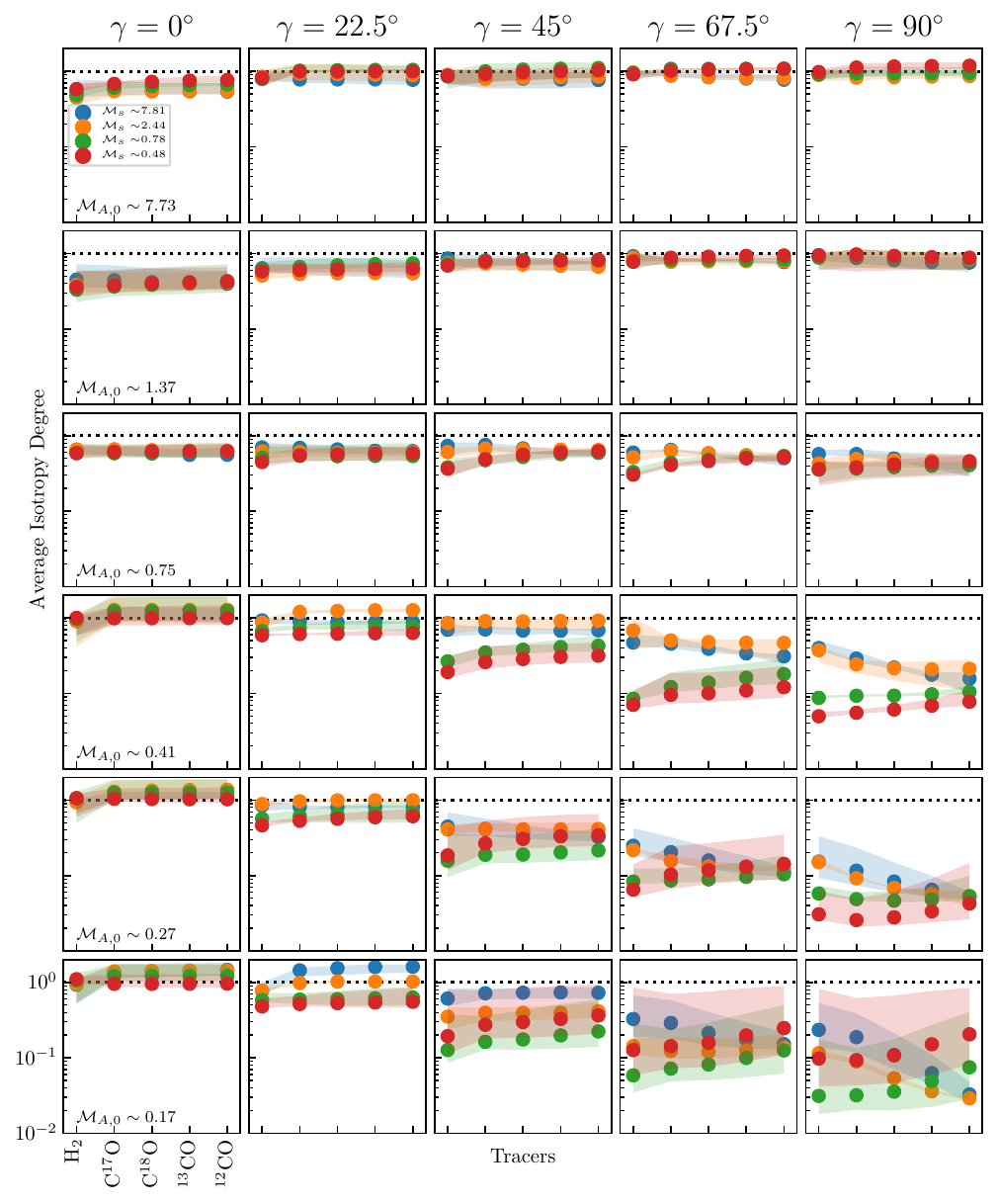}
	\caption{Average Isotropy Degree of Original Velocity for the different  tracers, with the same arrangement as Figure \ref{fig:AidI}.}
    \label{fig:AidOVC}
\end{figure*}

In Figure \ref{fig:AidOVC} we show the average isotropy degree from the velocity centroid maps (obtained with the original velocity field) with the same arrangement as the previous figure.
Similarly to the integrated intensity maps (Figure \ref{fig:AidI}) the isotropy degree changes for any given tracer from close to isotropic at smaller viewing angles to increasingly anisotropic for larger viewing angles and mean magnetic field magnitude.

Compared to the integrated intensity maps, the centroids are more anisotropic,
showing a significant anisotropy at lower viewing angles ($\gamma \geq 45\degr$) than in integrated intensity maps.
We can also observe a distinction between supersonic models ($\mathcal{M}_s>1$, in blue and orange) and subsonic models ($\mathcal{M}_s<1$, in green and red). The supersonic models, in general, are more isotropic than the subsonic ones.
For small viewing angles we there is only a slight dependence on the average isotropy degree with the Tracer used.
%Nevertheless, we can see that these same models are more anisotropic for thin tracers.
For viewing angles $\gamma \geq 45\degr$ and sub-Alfv\'enic turbulence (where the anisotropy is the largest) we do observe a trend with opacity for the supersonic models (as in the case of integrated intensity) where the anisotropy increases with optical depth. At the same time, for subsonic models there is only a slight dependence, and in some instance an increase of isotropy with optical depth. Models that are sub-sonic and at the same time super-Alvf\'enic have the largest uncertainty in the average isotropy degree, attributable to scale dependence in the range of scales used to measure the anisotropy.

%Compared to integrated intensity maps, centroids are more anisotropic, and such anisotropy does not vary significantly across tracers with different opacities. It is interesting to notice that the centroids are less affected by absorption, and how their anisotropy is remarkably similar across different tracers. The velocity centroids are more focused on the wings of the line, which is less affected by absorption.

%We have also included the analytical predictions in \citet{2017MNRAS.464.3617K} for the different  MHD modes. The comparison between the (optically thin) predictions and the models is reasonable (as reported in \citetalias{2020ApJ...901...11H}), and with marginal changes across the different tracers.

%\begin{figure*}[htp]
%    \centering
%	\includegraphics[scale=1]{fig6.pdf}
%	\caption{Average Isotropy Degree of Original Velocity Centroids vs. Alfv\'enic Mach number, with the same arrangement as Figure \ref{fig:AidI}. We include the analytic predictions in \citet{2017MNRAS.464.3617K} for the Alfvén mode (black line) and the slow mode (for high $\beta$ in orange and low $\beta$ in green)}
%    \label{fig:AidOVC}
%\end{figure*}

%%%%%%%%%%%%%%%%%%%%%%%%%%%%%%%%%%%%%%%%%%%%%%%%%%%%%
%%%%%%%%%%%%%%%%%%%%%%%%%%%%%%%%%%%%%%%%%%%%%%%%%%%%%

\section{Effect of turbulence strength on the depth probed by observations}
\label{sec:depth}

The anisotropy of turbulence in the volume sampled is determined by the anisotropy of the largest eddies present in the volume \citep{2002ApJ...564..291C}. However, absorption decreases the LOS extent of the volume sampled. The largest eddies, according to the theory of MHD turbulence (\citetalias{1995ApJ...438..763G}, see also book by \citealt{BeresnyakLazarian+2019}), are more isotropic. Their presence could mask the higher degree of the anisotropy of the small eddies.

One of the remarkable features we found in the previous analysis is the fact that the integrated intensity maps are very anisotropic for higher opacity lines.
We have attributed this behavior to the limited depth traced by optically thick lines, being unable to probe the entire \emph{cloud} they highlight the higher anisotropy of small scales. At the same optically thin lines which do probe the entire cloud include the contribution of larger and more isotropic eddies, averaging out some of the observed anisotropy.
%\esq{We should understand/discuss why subsonic centroids behave the way they do. Why they seem to ``saturate" the anisotropy degree with optical thickness, and even opposite in some instances, with anisotropy that decreases with opacity.}

The anisotropy in real space, for a given orientation (viewing angle) is determined by the magnetization (Alfv\'en Mach number), however as observed in the previous section, the sonic Mach number can have an impact on the resulting PPV anisotropy. As a measure of the size of scales probed by optical lines let us consider the distance traveled into the cloud until one reaches an optical depth of unity. If we repeat this at every position of the sky we are left with a surface of the $\tau=1$ boundary. In the case of a uniform medium and constant LOS velocity the resulting surface would be a plane aligned perpendicularly with the LOS. However in a turbulent environment the surface is highly corrugated due to velocity and density fluctuations.

\begin{figure*}[htp]
    \centering
	\includegraphics[scale=0.95]{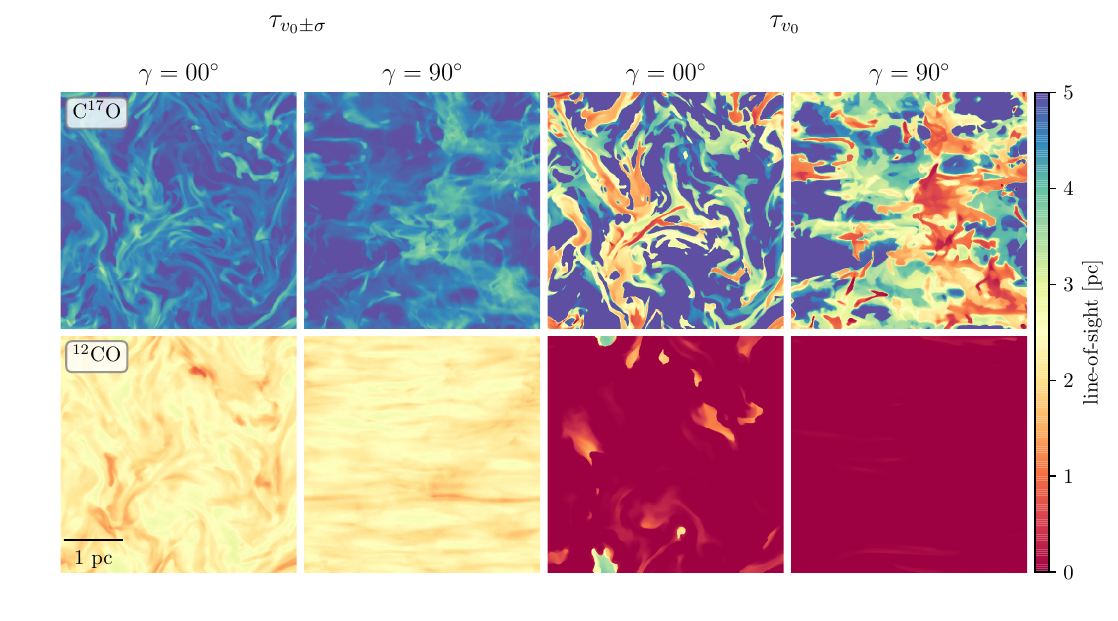}
	\caption{Distance (from the observed into the cloud) until  $\tau=1$ for the model M13 (a magnetized and supersonic model, $\mathcal{M}_A \sim 0.38$ and $\mathcal{M}_s \sim 7.62$) at $\pm \sigma$ from the center velocity channel (two first columns, $\tau_{v_0\pm \sigma}$) and in the center velocity channel (two last right columns, $\tau_{v_0}$). In rows, have results from $\mathrm{C^{17}O}$ maps at the top and $\mathrm{^{12}CO}$ at the bottom.
	At the top of each column we show the orientation considered ($\gamma=0\degr$, 1st and 3rd columns) and perpendicular to magnetic field ($\gamma=90\degr$, 2nd and 4th columns). The color-bar indicates the distance along the LOS in units of  parsec (the LOS extent of the cloud is 5 pc).}
    \label{fig:M13tau1}
\end{figure*}

We show in Figure \ref{fig:M13tau1} the distance traversed into the cloud until one reaches an optical depth of unity, for the extremes of optically thin $\mathrm{C^{17}O}$ and optically thick $\mathrm{^{12}CO}$ (top and bottom rows, respectively). Since the optical depth depends on the LOS velocity we include two cases, in the first two columns we present the average distance within a velocity range of $v_0 \pm \sigma_\mathrm{v}$ and in the two last columns the distance considering only the opacity at the line center ($v_0$). The average values of the optical depth are  also included in Table \ref{tab:models}. In the Figure we present the results obtained considering two viewing angles as indicated at the top of each column of plots.

We can see from the figure that the even for the thinnest CO line considered ($\mathrm{C^{17}O}$), the distance probed is less than the entire length of the cloud ($5~\mathrm{pc}$). If we consider the average distance considering velocities between $\pm \sigma_\mathrm{v}$ the distance probed is somewhere between $\sim 80\%$ and $100\%$ of the cloud, resulting in a low anisotropy in the maps. If only the opacity at the center of the line is considered we see that the $\tau=1$ surface is highly corrugated, with regions that traverse the entire cloud and patches that do not go beyond 1 pc ($20\%$ of the cloud thickness) but the anisotropy remains at a low level.
The reason of the extremely corrugated $\tau=1$ surface is the density fluctuations (e.g. shocks) in this highly supersonic model.
The distance probed by the optically thick $\mathrm{^{12}CO}$ line is significantly smaller, never reaching past $\sim50\%$ of the cloud thickness if we consider velocities between $\pm \sigma_\mathrm{v}$, and limited mostly to less than $\sim20\%$ of the cloud thickness in the center of the line. Thus resulting in a clear anisotropy for a LOS perpendicular to the mean magnetic field.

\begin{figure*}[htp]
    \centering
	\includegraphics[scale=0.95]{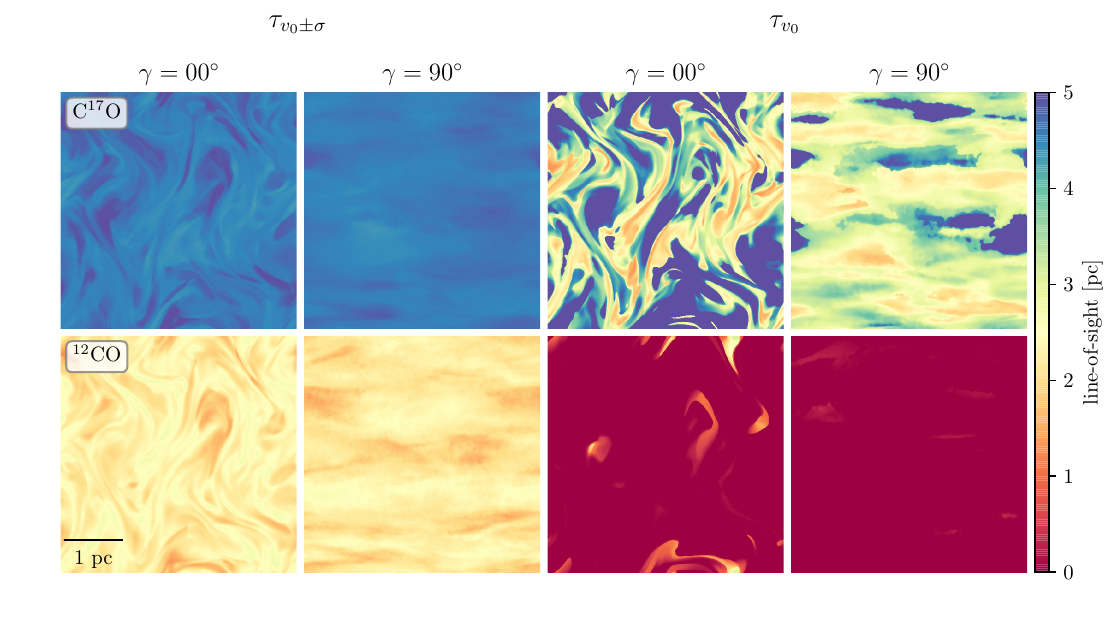}
	\caption{Distance until  $\tau=1$ for the model M16 (magnetized ans sub-sonic, $\mathcal{M}_A \sim 0.42$ and $\mathcal{M}_s \sim 0.59$). The figure has the same organization as Figure \ref{fig:M13tau1}}
    \label{fig:M16tau1}
\end{figure*}

In Figure in Figure \ref{fig:M16tau1}.  we illustrate the depth probed in model M16,  wich has a similar Alfv\'enic Mach number ($\mathcal{M}_A \sim 7.49$),  but a lower sonic Mach number ( and $\mathcal{M}_s \sim 0.53$).
We observe that in this subsonic case the $\tau=1$ surfaces are smoother  (all panels in the figure have a lower contrast in comparison with Fig. \ref{fig:M13tau1}); which translates in a lower level of large scale-scale mixing, and thus a more prominent anisotropy.

%To depict the behavior of all the models, we show in Figure \ref{fig:AidFVC} the maximum value of the optical depth for all the CO isotopologues. We can see from the figure that for subsonic turbulence there is very little dispersion and the opacity is small. And, naturally, as we increase the sonic Mach number the density fluctuations become stronger, and there is a larger dispersion,  and in general a larger maximum opacity which translates in a loss of anisotropy due to the mixture of scales.

\begin{figure*}[htp]
    \centering
	\includegraphics[scale=0.6]{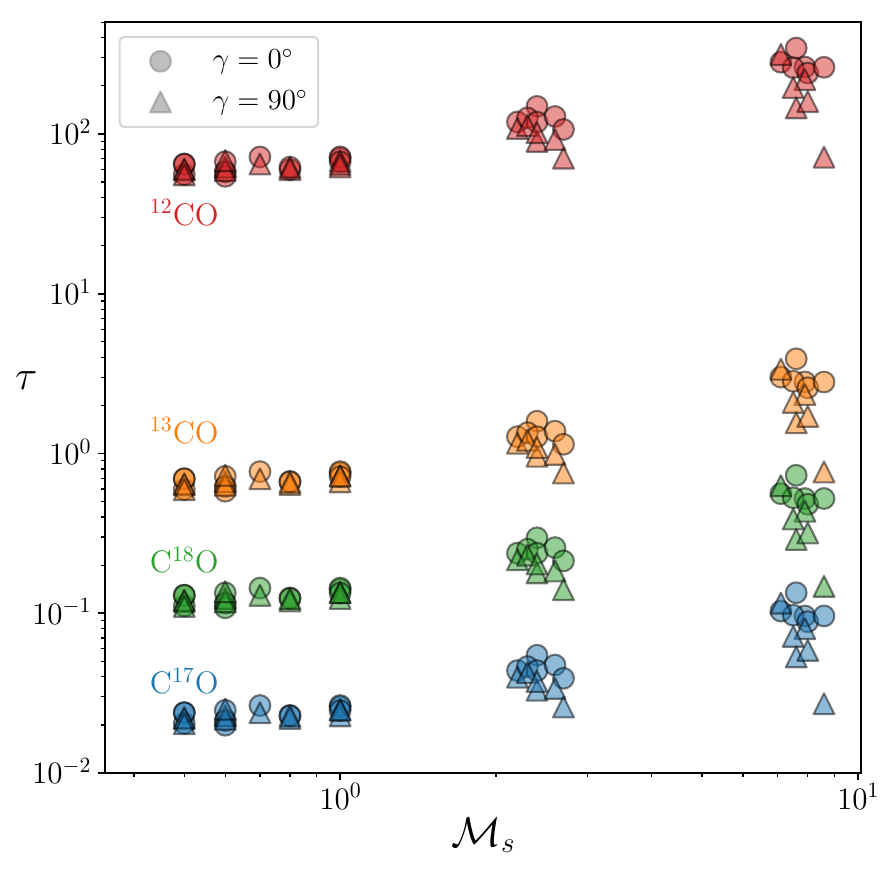}
	\caption{Maximum optical depth (averaged from $v=v_0 \pm \sigma_\mathrm{v}$) as a function of the sonic Mach number for all the models and different tracers. Grouped by colors are the different tracers  $\mathrm{^{12}CO}$ (red), $\mathrm{^{13}CO}$  (orange), $\mathrm{C^{18}O}$ (green), and $\mathrm{C^{17}O}$ (blue). In circles are the results for a LOS parallel to the mean magnetic field, and the triangles correspond to a LOS perpendicular to the mean magnetic field.}
    \label{fig:AidFVC}
\end{figure*}

To illustrate the behavior in all the models, we show in Figure \ref{fig:AidFVC}  the maximum value of the optical depth for all the CO isotopologues. We can see from the figure that there is very little dispersion of the models with different magnetization (each cluster of symbols correspond to models with different $\mathcal{M}_\mathrm{A}$ and similar $\mathcal{M}_\mathrm{s}$), for subsonic turbulence when the density fluctuations and scale mixing are low. Naturally, the density fluctuations become more substantial as we increase the sonic Mach number, which translates into a more significant dispersion and, in general, a considerably larger maximum opacity, which in turn yields to a loss of anisotropy due to the mixture of scales.

%\esq{David: I wonder if a better measure would be the dispersion of $\tau$ (it sould be a direct measure of the corrugation of the $\tau=1$ surface), and/or if we should present the same  results as a function of $M_A$ (to see if there is a trend between more and less magnetized moels).}

\section{Discussion}
\label{sec:disc}

Obtaining the directions of magnetic field as well as media magnetization using structure functions is gaining it momentum since the time of the introduction of the approach \citep{2002ASPC..276..182L, 2005ApJ...631..320E}. The recent developments include the the development of ways of accurate measurement of Alfv\'en Mach number using Structure Function Analysis \citep[SFA,][]{2021ApJ...910...88X,2021ApJ...912....2H}. A related avenue for recent research is connected to the Velocity Gradient Technique \citep[VGT,][]{2017ApJ...837L..24Y,2018ApJ...853...96L,2018ApJ...865...46L,2019ApJ...886...17H,2022MNRAS.513.3493H} that employs structure functions in the limit of minimal possible separation of correlating points \citep[see ][]{2020MNRAS.496.2868L,2020arXiv200207996L}.

Within the VGT, the effects of self-absorption were studied numerically in \citet{2019ApJ...873...16H} and this study adds to their analysis a better understanding of how different MHD turbulent modes are affected by the absorption effect. As for the SFA, our results show that the absorption does not decrease the ability of the technique to trace magnetic field.

The study of how the density statistics varies in the presence of self-absorption is also important. For sub-Alfvénic turbulence, density acts mostly as a passive scalar \citep[see ][]{BeresnyakLazarian+2019} and reflects the structure of the underlying turbulent velocity field. We note that this is true for solenoidally driven turbulence. Slow modes may be more critical for sub-Alfvenic turbulence. In this case, intensity gradients can trace the magnetic field the same way as the velocity gradients do \citep{2017ApJ...837L..24Y}. In supersonic turbulence, comparing the intensity and velocity gradients can reveal shocks \citep{2019ApJ...886...17H}. The latter in the low beta medium of molecular clouds can be associated with the steepened compressible part of the MHD turbulence, i.e. with the combination of slow and fast modes. This study clarifies how the properties of structure functions of intensity change in the presence of self-absorption.

%%%%%%%%%%%%%%%%%%%%%%%%%%%%%%%%%%%%%%%%%%%%%%%%%%%%%
%%%%%%%%%%%%%%%%%%%%%%%%%%%%%%%%%%%%%%%%%%%%%%%%%%%%%
\section{Summary}
\label{sec:sum}

We extend the work done in \citetalias{2020ApJ...901...11H} in which we address the anisotropy in optically thin synthetic spectroscopic observations.  In particular the anisotropy of structure-functions of maps of integrated intensity and of velocity centroids. In the present paper, we relax the assumption of an optically thin media. With this purpose, we perform in a post-processing step a radiation transfer code to obtain the emission of different CO isotopologues molecular lines, that range from optically thick to optically thin ($\mathrm{^{12}CO}$, $\mathrm{^{13}CO}$, $\mathrm{C^{18}O}$, and $\mathrm{C^{17}O}$).

The maps of integrated intensity are found to be isotropic for super-Alfv\'enic turbulence irregardless of the tracer considered (e.g. see first two rows of Fig. \ref{fig:AidI}). For sub-Alfv\'enic turbulence (third to fifth rows in Fig. \ref{fig:AidI}) the models exhibit anisotropy aligned with the direction of the mean magnetic field, i.e. the isotropy degree less than unity (there are a handful cases in which the anisotropy is not aligned with the mean magnetic field, but within the error bars they can be considered isotropic).
In the highest magnetization models we found a dependence with the tracer used, and the sonic Mach number at large viewing angles. There is a lower level of anisotropy for the optically thin H$_2$ and $\mathrm{C^{17}O}$ maps. For the rest of the tracers we found a more pronounced anisotropy; but whilst for $\mathcal{M}_\mathrm{s}<1$ they seem to saturate at level of anisotropy in the $\mathrm{C^{18}O}$ map, for $\mathcal{M}_\mathrm{s}>1$ the anisotropy keeps increasing monotonically with opacity.

We find that the anisotropy of velocity centroids in the different CO tracers is remarkably similar to the optically thin case results. That is, iso-contours of the structure-function of the velocity centroids align with the direction of the mean magnetic field projected into the plane of the sky, with a higher degree of elongation for models with a higher degree of magnetization. Only small differences are noticeable, with a higher anisotropy for optically thick lines.

Also, as done in \citetalias{2020ApJ...901...11H} we decompose the velocity field into the contribution to the different MHD (Alfv\'en, slow and fast) modes and the results obtained for optically thin lines hold. The fast mode has a marginal contribution across our models.
The Alfv\'en mode dominates where the line of sight has a large angle ($\gtrsim 45\degr$) with the mean magnetic field, and the slow mode dominates for small viewing angles. In general, the observed anisotropy  can be attributed  to the Alfv\'en mode (see Figs. \ref{fig:OVC} and \ref{fig:MHDmodes}). However, we must point out that our assumption of incompressible driving, and the lack of self-gravity could underestimate the contribution in the fast-mode.

In contrast with the results obtained for optically thin lines, we found that for optically thick tracers
%(such as $\mathrm{^{12}CO}$ and to a lesser degree in  $\mathrm{^{13}CO}$)
the integrated intensity maps become remarkably anisotropic and their structure-function iso-contours align with the mean magnetic field on the plane of the sky. We attribute this to the limited depth reached into the cloud by these tracers, limiting the emission to small scales, which are known to be more anisotropic (\citetalias{1995ApJ...438..763G}, see Figure \ref{fig:CD}).

An important result of this paper is that the degree of anisotropy measured in velocity centroids does not change considerably in tracers within a large range of opacities. This is encouraging because it means that most earlier studies of anisotropy in centroids made with different tracers do not require additional corrections. At the same time, the anisotropy of integrated intensity does require more careful analysis.

%We showed how the strength of the turbulence (sonic Mach number) modifies the distance that can be probed by optically thick lines and thus the resulting anisotropy \dvd{(see the second row in Figures \ref{fig:M13tau1} and \ref{fig:M4tau1})}.
We showed how the strength of the turbulence (sonic Mach number) modifies the distance that optically thick lines can probe and thus the resulting anisotropy (see the second row in Figures \ref{fig:M13tau1} and \ref{fig:M16tau1}).
Higher sonic Mach numbers create regions of high density and also open low-density canals that result in a highly corrugated $\tau =1 $ surface, thus reducing the anisotropy by mixing several scales (see the first row of Figure \ref{fig:M13tau1}). Conversely, for low sonic Mach numbers, the density contrast decreases, yielding to a smoother $\tau =1 $ surface and a less mixing of scales, preserving better the anisotropy (see the first row of Figure \ref{fig:M16tau1}).

%%%%%%%%%%%%%%%%%%%%%%%%%%%%%%%%%%%%%%%%%%%%%%%%%%%%%
%%%%%%%%%%%%%%%%%%%%%%%%%%%%%%%%%%%%%%%%%%%%%%%%%%%%%
\begin{acknowledgements}

DHP, AE, and PFV acknowledge financial support from PAPIIT-UNAM grants IG100422 and IN113522. We thank Enrique Palacios for maintaining the Diable cluster, where the simulations were performed.

\end{acknowledgements}

%%%%%%%%%%%%%%%%%%%%%%%%%%%%%%%%%%%%%%%%%%%%%%%%%%
%%%%%%%%%%%%%%%%%%%% APPENDIX %%%%%%%%%%%%%%%%%%
\appendix

\section{Average Isotropy Degree for MHD modes}

In section \ref{sec:ID} we described the average isotropy degree for both the integrated intensity and the original velocity centroids. Here we include the average degree of isotropy for the Alfvén, slow, and fast MHD modes.

\begin{figure*}[htp]
    \centering
	\includegraphics[scale=1]{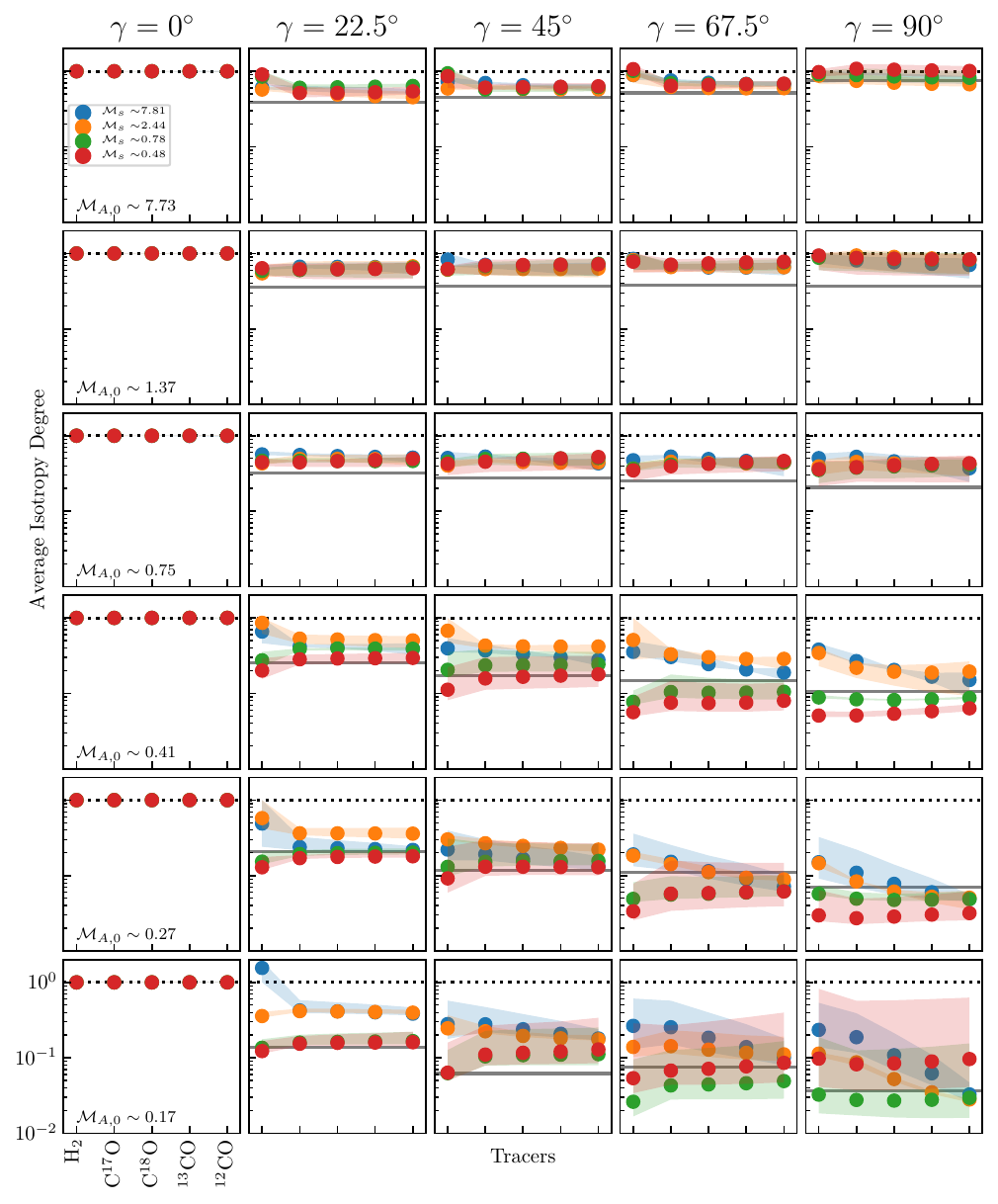}
	\caption{Average Isotropy Degree of Alfvén mode Velocity Centroids for the different tracers, with the same arrangement as Figure \ref{fig:AidI}. The solid black lines correspond to the analytic predictions in \citet{2017MNRAS.464.3617K} for the Alfvén mode}
    \label{fig:IDalfven}
\end{figure*}

Figure \ref{fig:IDalfven} shows the average isotropy degree of the Alfvén MHD mode velocity centroids vs. the different tracers. In this figure, we have the same arrangement as in Figure \ref{fig:AidOVC} where we also include the analytical prediction of \citet{2017MNRAS.464.3617K} (the solid black lines). We notice that all models remain isotropic for the viewing angle $\gamma = 0\degr$ since the Alfvén waves displace the material perpendicularly to the magnetic field if the magnetic field is aligned with the LOS there is only a residual signal. We must note that the predictions from \citet{2017MNRAS.464.3617K} are not valid for this orientation, being only a formal limit of the anisotropy but a zero amplitude. A similar situation is seen with the slow modes, where the $\gamma = 90\degr$ is also a zero amplitude prediction. We also noticed two slopes, a positive one for the subsonic models and a negative one for the supersonic ones, as we saw in Figure \ref{fig:AidOVC}. This is more noticeable for models with higher magnetization ($\mathcal{M}_A<1$) and with  $\gamma \gtrsim 22.5\degr$.

On the other hand, we also note that for angles of view $\gamma \gtrsim 45\degr$, we have a distribution of the models very similar to the average isotropy degree for the original velocity centroids (see Figure \ref{fig:AidOVC}). This same distribution of the models had already been observed and explained in \citetalias{2020ApJ...901...11H}.

\begin{figure*}[htp]
    \centering
	\includegraphics[scale=1]{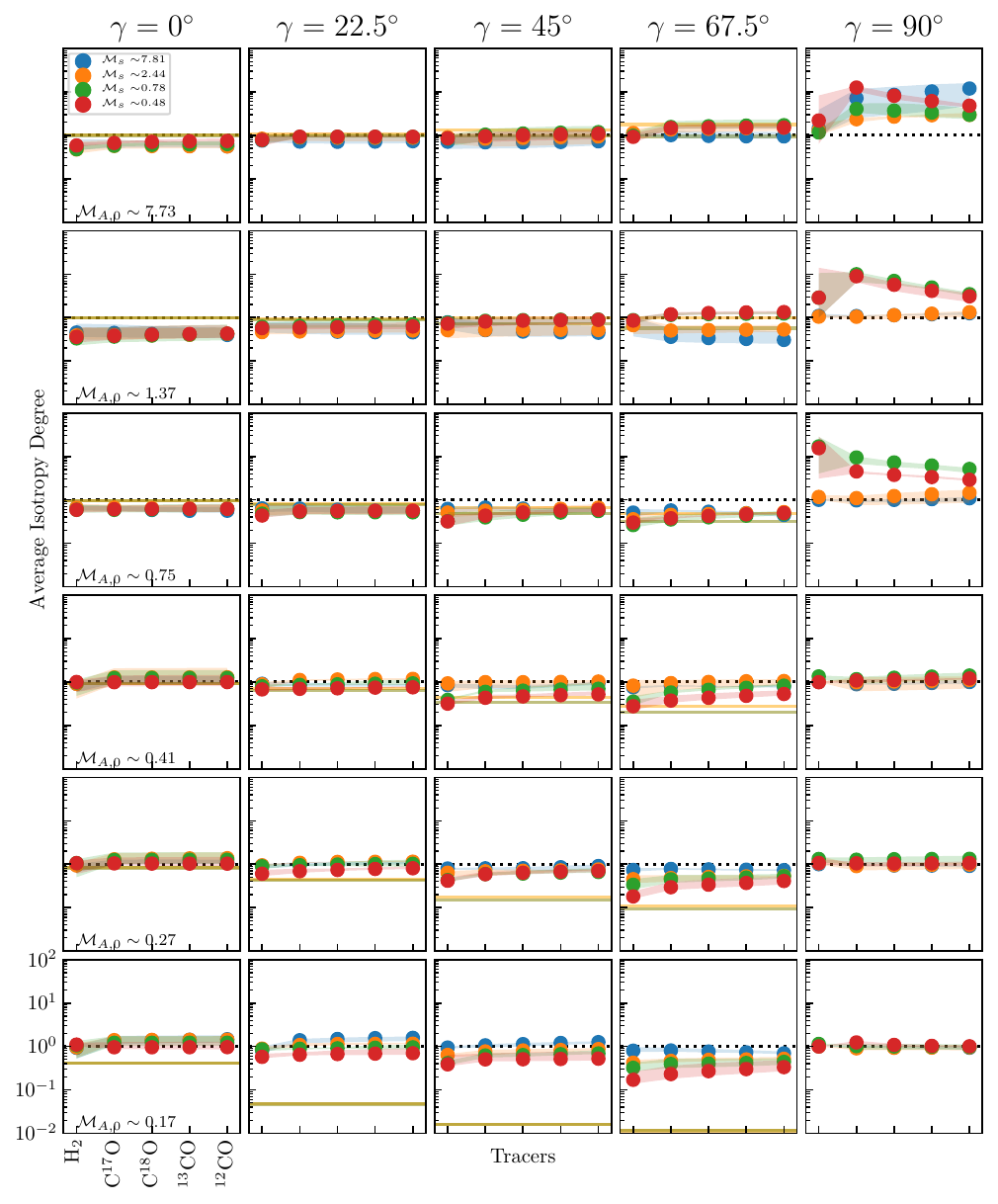}
	\caption{Average Isotropy Degree of Slow mode Velocity Centroids for the different tracers, with the same arrangement as Figure \ref{fig:AidI}. The solid lines correspond to the the analytic predictions in \citet{2017MNRAS.464.3617K} for the slow mode (for high $\beta$ in orange and low $\beta$ in green)}
    \label{fig:IDslow}
\end{figure*}

Figure \ref{fig:IDslow} show the average isotropy degree of the slow MHD mode velocity centroids vs. the tracers. It has the same arrangement as Figure \ref{fig:AidOVC} and now include the analytical prediction of \citet{2017MNRAS.464.3617K}, for high $\beta$ in orange and low $\beta$ in green.
In this Figure, we observe that the less magnetized models are more isotropic. If we compare them with Figure \ref{fig:AidOVC}, we can notice that the models with $\gamma \lesssim 22.5\degr$ behave very similarly. For $\gamma = 90\degr$ slow MHD contribute only marginally to the velocity isotropy. This was also observed in \citetalias{2020ApJ...901...11H}.

\begin{figure*}[htp]
    \centering
	\includegraphics[scale=1]{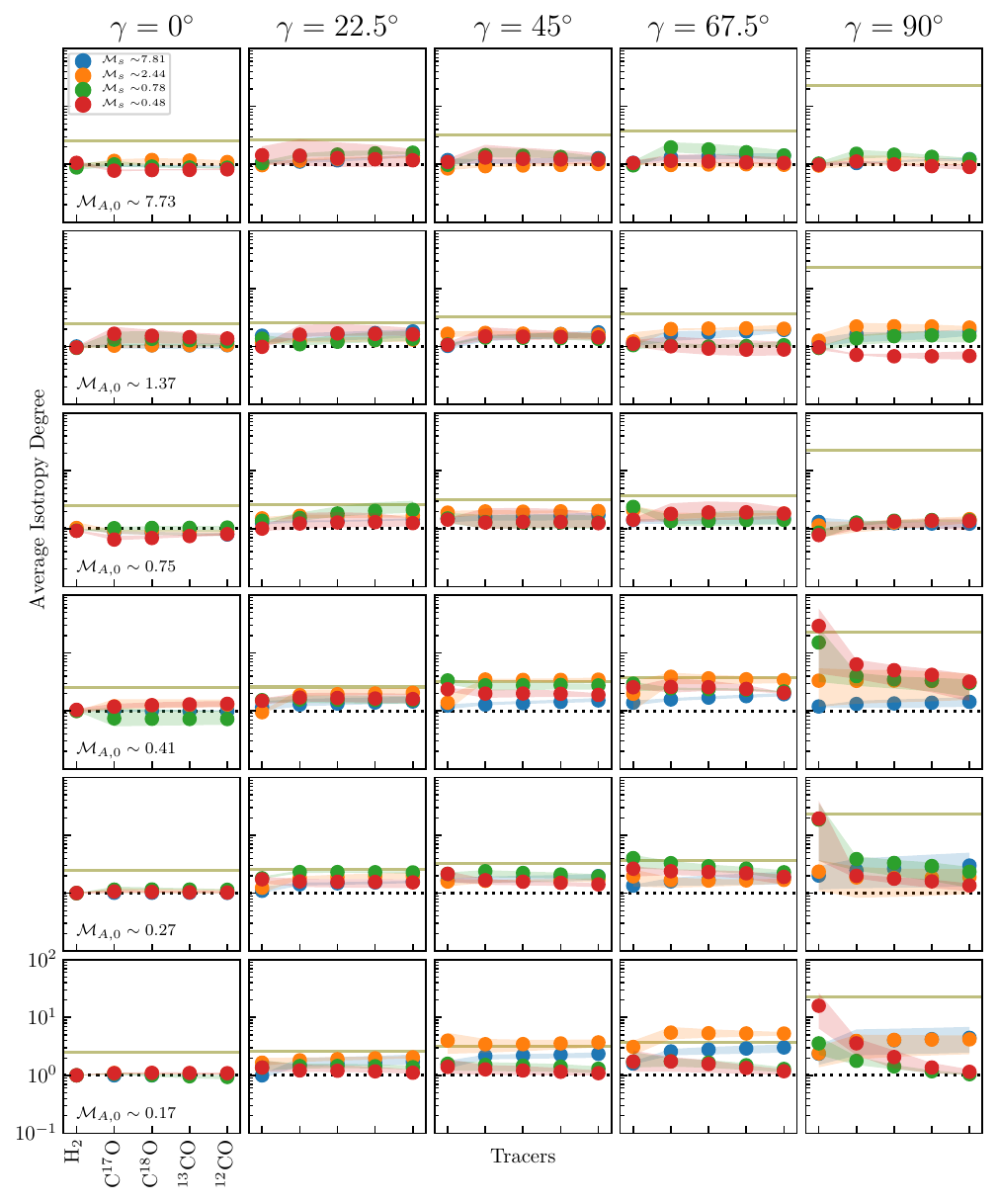}
	\caption{Average Isotropy Degree of Fast mode Velocity Centroids for the different  tracers, with the same arrangement as Figure \ref{fig:AidI}}
    \label{fig:IDfast}
\end{figure*}

Figure \ref{fig:AidFVC} show the average isotropy degree of the fast MHD mode velocity centroids vs. the tracers. As the previous Figures, this has the same arrangement as Figure \ref{fig:AidOVC}. We include these for completeness, as their contribution is negligible in our models. It would be of interest to see how ther contribution change with a non purely solenoidal turbulence driving, of self gravity, both of which should increase the role of the fast megneto-sonic mode.

%\bibliography{Master}{}
%\bibliographystyle{aasjournal}

\end{document}